\documentclass[onecolumn]{elsart3p}
\usepackage{ifpdf}
\usepackage[usenames]{color}
\usepackage{graphicx}
\usepackage{amssymb}
\usepackage{amsmath}

\journal{Computer Physics Communications}

\begin{document}

\begin{frontmatter}

\title{{Fortran programs for the time-dependent
Gross-Pitaevskii equation in a fully anisotropic trap}}


\author[bdu,ift]{P. Muruganandam} and
\ead{anand@cnld.bdu.ac.in}
\author[ift]{S. K. Adhikari\corauthref{cor1}}%
\ead{adhikari@ift.unesp.br}
\address[bdu]{School of Physics, Bharathidasan University, Palkalaiperur Campus,
Tiruchirappalli -- 620024, Tamil Nadu, India}
\address[ift]{Instituto de F\'{\i}sica Te\'{o}rica, UNESP -- S\~{a}o
Paulo State University, Barra Funda, 01.140-70 S\~{a}o Paulo, S\~{a}o Paulo, Brazil}
\corauth[cor1]{Corresponding author; address: Instituto de F\'{\i}sica Te\'{o}rica, Rua Pamplona 145, 01405-900 S\~ao Paulo, SP, Brazil; telephone: +55 11
3177 9071}

\begin{abstract} 

Here we develop simple numerical algorithms for both stationary and
non-stationary solutions of the time-dependent Gross-Pitaevskii (GP) equation
describing the properties of Bose-Einstein condensates at ultra low
temperatures. In particular, we consider  algorithms involving real- and
imaginary-time propagation based on a split-step Crank-Nicolson method. In a
one-space-variable form of the GP equation we consider the one-dimensional, 
two-dimensional circularly-symmetric, and the three-dimensional
spherically-symmetric harmonic-oscillator traps. In the 
two-space-variable form we consider the GP
equation in two-dimensional anisotropic  and three-dimensional axially-symmetric
traps. The fully-anisotropic three-dimensional GP equation is also considered.
Numerical results for the chemical potential and root-mean-square size of
stationary states are reported using imaginary-time propagation programs for all
the cases and compared with previously obtained results. Also presented are
numerical results of non-stationary oscillation for different trap symmetries
using real-time propagation programs. A set of convenient working codes
developed in Fortran 77  are also provided for all these cases (twelve programs
in all). In the case of two or three space variables, { Fortran
90/95 versions provide some simplification over the Fortran 77 programs}, and
these programs are also included (six programs in all). 

\end{abstract}


\end{frontmatter}

\maketitle

{\bf Program summary (1)}

Title of program: imagtime1d.F

Title of electronic file: imagtime1d.tar.gz

Catalogue identifier:

Program summary URL: 

Program obtainable from: CPC Program Library, Queen's University of 
Belfast, N. Ireland

Distribution format: tar.gz

Computers: PC/Linux, workstation/UNIX

Maximum Ram Memory: 1 GByte

Programming language used: Fortran 77




Typical running time: Minutes on a medium PC

Unusual features: None

Nature of physical problem: This program is designed to solve the
time-dependent Gross-Pitaevskii nonlinear partial differential equation
in one space dimension with a harmonic trap. The Gross-Pitaevskii
equation describes the properties of a dilute trapped Bose-Einstein
condensate.

Method of solution: The time-dependent Gross-Pitaevskii equation is 
solved by the split-step Crank-Nicolson method by discretizing in space 
and time. The discretized equation is then solved by propagation in 
imaginary time over small time steps.  The method yields the solution of 
stationary problems.

{\bf Program summary (2)}

Title of program: imagtimecir.F

Title of electronic file: imagtimecir.tar.gz

Catalogue identifier:

Program summary URL: 

Program obtainable from: CPC Program Library, Queen's University of 
Belfast, N. Ireland

Distribution format: tar.gz

Computers: PC/Linux, workstation/UNIX

Maximum Ram Memory: 1 GByte

Programming language used: Fortran 77




Typical running time: Minutes on a medium PC

Unusual features: None

Nature of physical problem: This program is designed to solve the
time-dependent Gross-Pitaevskii nonlinear partial differential equation
in two space dimensions with a circularly-symmetric trap. The
Gross-Pitaevskii equation describes the properties of a dilute trapped
Bose-Einstein condensate.

Method of solution: The time-dependent Gross-Pitaevskii equation is 
solved by the split-step Crank-Nicolson method by discretizing in space 
and time. The discretized equation is then solved by propagation in 
imaginary time over small time steps.  The method yields the solution of 
stationary problems.

{\bf Program summary (3)}

Title of program: imagtimesph.F

Title of electronic file: imagtimesph.tar.gz

Catalogue identifier:

Program summary URL: 

Program obtainable from: CPC Program Library, Queen's University of 
Belfast, N. Ireland

Distribution format: tar.gz

Computers: PC/Linux, workstation/UNIX

Maximum Ram Memory: 1 GByte

Programming language used: Fortran 77




Typical running time: Minutes on a medium PC

Unusual features: None

Nature of physical problem: This program is designed to solve the
time-dependent Gross-Pitaevskii nonlinear partial differential equation
in three space dimensions with a spherically-symmetric trap. The
Gross-Pitaevskii equation describes the properties of a dilute trapped
Bose-Einstein condensate.

Method of solution: The time-dependent Gross-Pitaevskii equation is 
solved by the split-step Crank-Nicolson method by discretizing in space 
and time. The discretized equation is then solved by propagation in 
imaginary time over small time steps.  The method yields the solution of 
stationary problems.

{\bf Program summary (4)}

Title of program: realtime1d.F

Title of electronic file: realtime1d.tar.gz

Catalogue identifier:

Program summary URL: 

Program obtainable from: CPC Program Library, Queen's University of 
Belfast, N. Ireland

Distribution format: tar.gz

Computers: PC/Linux, workstation/UNIX

Maximum Ram Memory: 2 GByte

Programming language used: Fortran 77




Typical running time: Minutes on a medium PC

Unusual features: None

Nature of physical problem: This program is designed to solve the
time-dependent Gross-Pitaevskii nonlinear partial differential equation
in one space dimension with a harmonic trap. The Gross-Pitaevskii
equation describes the properties of a dilute trapped Bose-Einstein
condensate.

Method of solution: The time-dependent Gross-Pitaevskii equation is 
solved by the split-step Crank-Nicolson method by discretizing in space 
and time. The discretized equation is then solved by propagation in 
real time over small time steps.  The method yields the solution of 
stationary and non-stationary problems.

{\bf Program summary (5)}

Title of program: realtimecir.F

Title of electronic file: realtimecir.tar.gz

Catalogue identifier:

Program summary URL: 

Program obtainable from: CPC Program Library, Queen's University of 
Belfast, N. Ireland

Distribution format: tar.gz

Computers: PC/Linux, workstation/UNIX

Maximum Ram Memory: 2 GByte

Programming language used: Fortran 77




Typical running time: Minutes on a medium PC

Unusual features: None

Nature of physical problem: This program is designed to solve the
time-dependent Gross-Pitaevskii nonlinear partial differential equation
in two space dimensions with a circularly-symmetric trap. The
Gross-Pitaevskii equation describes the properties of a dilute trapped
Bose-Einstein condensate.

Method of solution: The time-dependent Gross-Pitaevskii equation is 
solved by the split-step Crank-Nicolson method by discretizing in space 
and time. The discretized equation is then solved by propagation in 
real time over small time steps.  The method yields the solution of 
stationary and non-stationary problems.

{\bf Program summary (6)}

Title of program: realtimesph.F

Title of electronic file: realtimesph.tar.gz

Catalogue identifier:

Program summary URL: 

Program obtainable from: CPC Program Library, Queen's University of 
Belfast, N. Ireland

Distribution format: tar.gz

Computers: PC/Linux, workstation/UNIX

Maximum Ram Memory: 2 GByte

Programming language used: Fortran 77




Typical running time: Minutes on a medium PC

Unusual features: None

Nature of physical problem: This program is designed to solve the
time-dependent Gross-Pitaevskii nonlinear partial differential equation
in three space dimensions with a spherically-symmetric trap. The
Gross-Pitaevskii equation describes the properties of a dilute trapped
Bose-Einstein condensate.

Method of solution: The time-dependent Gross-Pitaevskii equation is 
solved by the split-step Crank-Nicolson method by discretizing in space 
and time. The discretized equation is then solved by propagation in 
real time over small time steps.  The method yields the solution of 
stationary and non-stationary problems.

{\bf Program summary (7)}

Title of programs: imagtimeaxial.F  and imagtimeaxial.f90 

Title of electronic file: imagtimeaxial.tar.gz

Catalogue identifier:

Program summary URL: 

Program obtainable from: CPC Program Library, Queen's University of 
Belfast, N. Ireland

Distribution format: tar.gz

Computers: PC/Linux, workstation/UNIX

Maximum Ram Memory: 2 GByte

Programming language used: Fortran 77 and Fortran 90




Typical running time: Few hours on a medium PC

Unusual features: None

Nature of physical problem: This program is designed to solve the
time-dependent Gross-Pitaevskii nonlinear partial differential equation
in three space dimensions with an axially-symmetric trap. The
Gross-Pitaevskii equation describes the properties of a dilute trapped
Bose-Einstein condensate.

Method of solution: The time-dependent Gross-Pitaevskii equation is 
solved by the split-step Crank-Nicolson method by discretizing in space 
and time. The discretized equation is then solved by propagation in 
imaginary time over small time steps.  The method yields the solution of 
stationary problems.

{\bf Program summary (8)}

Title of program: imagtime2d.F and imagtime2d.f90

Title of electronic file: imagtime2d.tar.gz

Catalogue identifier:

Program summary URL: 

Program obtainable from: CPC Program Library, Queen's University of 
Belfast, N. Ireland

Distribution format: tar.gz

Computers: PC/Linux, workstation/UNIX

Maximum Ram Memory: 2 GByte

Programming language used: Fortran 77 and Fortran 90




Typical running time: Few hours on a medium PC

Unusual features: None

Nature of physical problem: This program is designed to solve the
time-dependent Gross-Pitaevskii nonlinear partial differential equation
in two space dimensions with an anisotropic trap. The Gross-Pitaevskii
equation describes the properties of a dilute trapped Bose-Einstein
condensate.

Method of solution: The time-dependent Gross-Pitaevskii equation is 
solved by the split-step Crank-Nicolson method by discretizing in space 
and time. The discretized equation is then solved by propagation in 
imaginary time over small time steps.  The method yields the solution of 
stationary problems.

{\bf Program summary (9)}

Title of program: realtimeaxial.F and realtimeaxial.f90

Title of electronic file: realtimeaxial.tar.gz

Catalogue identifier:

Program summary URL: 

Program obtainable from: CPC Program Library, Queen's University of 
Belfast, N. Ireland

Distribution format: tar.gz

Computers: PC/Linux, workstation/UNIX

Maximum Ram Memory: 4 GByte

Programming language used: Fortran 77 and Fortran 90




Typical running time: Hours on a medium PC

Unusual features: None

Nature of physical problem: This program is designed to solve the
time-dependent Gross-Pitaevskii nonlinear partial differential equation
in three space dimensions with an axially-symmetric trap. The
Gross-Pitaevskii equation describes the properties of a dilute trapped
Bose-Einstein condensate.

Method of solution: The time-dependent Gross-Pitaevskii equation is 
solved by the split-step Crank-Nicolson method by discretizing in space 
and time. The discretized equation is then solved by propagation in 
real time over small time steps.  The method yields the solution of 
stationary and non-stationary problems.

{\bf Program summary (10)}

Title of program: realtime2d.F and realtime2d.f90

Title of electronic file: realtime2d.tar.gz

Catalogue identifier:

Program summary URL: 

Program obtainable from: CPC Program Library, Queen's University of 
Belfast, N. Ireland

Distribution format: tar.gz

Computers: PC/Linux, workstation/UNIX

Maximum Ram Memory: 4 GByte

Programming language used: Fortran 77 and Fortran 90




Typical running time: Hours on a medium PC

Unusual features: None

Nature of physical problem: This program is designed to solve the
time-dependent Gross-Pitaevskii nonlinear partial differential equation
in two space dimensions with an anisotropic trap. The Gross-Pitaevskii
equation describes the properties of a dilute trapped Bose-Einstein
condensate.

Method of solution: The time-dependent Gross-Pitaevskii equation is 
solved by the split-step Crank-Nicolson method by discretizing in space 
and time. The discretized equation is then solved by propagation in 
real time over small time steps.  The method yields the solution of 
stationary and non-stationary problems.

{\bf Program summary (11)}

Title of program: imagtime3d.F and imagtime3d.f90

Title of electronic file: imagtime3d.tar.gz

Catalogue identifier:

Program summary URL: 

Program obtainable from: CPC Program Library, Queen's University of 
Belfast, N. Ireland

Distribution format: tar.gz

Computers: PC/Linux, workstation/UNIX

Maximum Ram Memory: 4 GByte

Programming language used: Fortran 77 and Fortran 90




Typical running time: Few days on a medium PC

Unusual features: None

Nature of physical problem: This program is designed to solve the
time-dependent Gross-Pitaevskii nonlinear partial differential equation
in three space dimensions with an anisotropic trap. The Gross-Pitaevskii
equation describes the properties of a dilute trapped Bose-Einstein
condensate.

Method of solution: The time-dependent Gross-Pitaevskii equation is 
solved by the split-step Crank-Nicolson method by discretizing in space 
and time. The discretized equation is then solved by propagation in 
imaginary time over small time steps.  The method yields the solution of 
stationary problems.

{\bf Program summary (12)}

Title of program: realtime3d.F and realtime3d.f90

Title of electronic file: realtime3d.tar.gz

Catalogue identifier:

Program summary URL: 

Program obtainable from: CPC Program Library, Queen's University of 
Belfast, N. Ireland

Distribution format: tar.gz

Computers: PC/Linux, workstation/UNIX

Maximum Ram Memory: 8 GByte

Programming language used: Fortran 77 and Fortran 90




Typical running time: Days on a medium PC

Unusual features: None

Nature of physical problem: This program is designed to solve the
time-dependent Gross-Pitaevskii nonlinear partial differential equation
in three space dimensions with an anisotropic trap. The Gross-Pitaevskii
equation describes the properties of a dilute trapped Bose-Einstein
condensate.

Method of solution: The time-dependent Gross-Pitaevskii equation is
solved by the split-step Crank-Nicolson method by discretizing in space
and time. The discretized equation is then solved by propagation in real
time over small time steps.  The method yields the solution of
stationary and non-stationary problems.

\section{Introduction}

After a successful experimental detection of Bose-Einstein condensates (BEC) of
dilute trapped bosonic alkali-metal atoms $^7$Li, $^{23}$Na, and $^{87}$Rb
\cite{review,books} at ultra-low temperatures, there have been intense
theoretical activities in studying properties of the condensate using the
time-dependent mean-field Gross-Pitaevskii (GP) equation under different trap
symmetries. Among many possibilities, the following traps have been used in
various studies: three-dimensional (3D) spherically-symmetric, axially-symmetric
and anisotropic harmonic traps, two-dimensional (2D) circularly-symmetric and
anisotropic harmonic traps, and one-dimensional (1D) harmonic trap. The
inter-atomic interaction leads to a nonlinear term in the GP equation, which
complicates its accurate numerical solution, specially for a large nonlinearity.
The nonlinearity is large for a fixed harmonic trap when either the number of
atoms in the condensate or the atomic scattering length is large and this is
indeed so under  many experimental conditions. Special care is needed 
for the
solution of the time-dependent GP equation with large nonlinearity and there has
been an extensive literature on this topic
\cite{Tiwari_Shukla,Bao_Tang,Schneider_Feder,chio1,chio2,chang,num1,num2,num3,num4,num5,num6,num7,num8,num9,num10,num11,num12,num13,num14,num15,num16,num17,num18,num19,num20,num21,num22,num23,num24,num25,num26,num27,num28,num29,num30,num31,num32,num33,num34,num35,aq,xyz1,burnett,holland,baer}.

The time-dependent GP equation is a partial differential equation in space and
time variables involving first-order time and second-order space derivatives
together with a harmonic and a nonlinear potential term, and has the structure
of a nonlinear Schr\"odinger equation with a harmonic trap. { One
commonly used} procedure for the solution of the time-dependent GP equation makes
use of a discretization of this equation in space and time  and subsequent
integration and time propagation of the discretized equation. From a knowledge
of the solution of this equation at a specific time, this procedure finds the
solution after a small time step by solving the discretized equation.
{A commonly used} discretization scheme for the GP equation is the
semi-implicit Crank-Nicolson discretization scheme \cite{koonin,ames,dtray}
which has certain advantages and will be used in this work.

In the simplest one-space-variable form of the GP  equation, the solution
algorithm is executed in two steps. In the first step, using a known initial
solution, an intermediate solution  after a small interval of time $\Delta$ is
found   neglecting the harmonic and  nonlinear potential terms. The effect of
the potential terms is then included by a first-order time integration to obtain
the final solution after time $\Delta$. In case of two or three spatial
variables, the space derivatives are dealt with in two or three steps and the
effect of the potential terms are included next. As the time evolution is
executed  in different steps it is called a split-step real-time propagation
method. This method is equally applicable to stationary ground and 
excited states as well as non-stationary states, although in this paper 
we do not consider stationary excited states. 
The virtues of the semi-implicit Crank-Nicolson scheme
\cite{koonin,ames,dtray} are that it is  unconditionally stable and preserves the
normalization  of  the solution under real-time propagation. A simpler 
and
efficient variant of the {scheme} called the split-step
imaginary-time propagation method obtained by replacing the time variable by an
imaginary time is also considered. (The GP equation involves complex
{variables}. However, after replacing the time variable by an
imaginary time the resultant partial differential equation is real, and hence
the imaginary-time propagation method involves real {variables}
only. {This trick
leads to  an 
imaginary-time operator which results in exponential decay of all states 
relative to the   ground state and   can then be applied to any 
initial 
trial wave function to compute an approximation to the actual ground 
state rather accurately. We shall use imaginary-time propagation to 
compute the ground state in this paper.)} The split-step 
imaginary-time propagation method involving real variables
 yields very precise result at low computational cost (CPU time) and is very
appropriate for the solution of  stationary problems involving the 
ground state. The split-step 
real-time propagation method uses complex {quantities} and yields
less precise results for stationary problems; however, they are appropriate for
the study of non-equilibrium dynamics in addition to stationary problems 
involving excited states also.

Most of the previous studies
\cite{Tiwari_Shukla,Bao_Tang,Schneider_Feder,chang,num1,num4,num10,num12,num13,num16,num17,num25,num26,num27,num31,xyz1,burnett}
on the numerical solution of the GP equation are confined to a consideration of
stationary states only. Some used specifically the imaginary-time propagation
method \cite{chio1,num23,aq,xyz1}. There are few studies
\cite{num22,num29,num30,num33} for the numerical solution of the time-dependent
GP equation using  the Crank-Nicolson method \cite{koonin,ames,dtray}. Other
methods for numerical solution of the time-dependent GP equation have also
appeared in the literature
\cite{chio2,num6,num8,num15,num18,num19,num20,num21,num24,num34,num35,holland,baer}.
These time-dependent methods can be used for studying non-equilibrium dynamics
of the condensate involving non-stationary states.

The purpose of the present paper is to develop {a simple and
efficient} algorithm for the numerical solution of the GP equation using  time
propagation together with the semi-implicit Crank-Nicolson discretization
scheme \cite{koonin,ames,dtray} specially useful to newcomers in this field
interested in obtaining a numerical solution of the time-dependent GP equation.
Easy-to-use Fortran 77  programs for different  trap symmetries with adequate
explanation are also included. In case of two and three space variables, Fortran
90/95 programs are more compact in nature and these programs are also included.
We include programs using both real- and imaginary-time propagation. For
stationary ground states the imaginary-time method has a much quicker 
convergence rate
compared to the real-time method and should be used for the calculation of
chemical potential, energy  and root-mean-square (rms) sizes. We calculate the chemical
potential and rms sizes of the condensate for stationary problems and compare
these results with those previously obtained by other workers for different trap
symmetries. These results can be easily calculated in a decent PC using the
Fortran programs provided. In addition to the results for the stationary states,
the real-time propagation routines can also be used to study the 
non-stationary
transitions,  as in  collapse dynamics \cite{ska5} and non-equilibrium
oscillation \cite{num22}.

{This paper is organized as follows.} In Sec. \ref{gpe} we present
the GP equations with different traps that  we consider in this paper, e.g., the
3D spherically-symmetric, 2D circularly-symmetric  and 1D harmonic traps
involving one space variable, the anisotropic 2D  and axially-symmetric 3D
harmonic traps in two space variables and the fully anisotropic 3D harmonic trap
in three space variables. In Sec. \ref{CN1D} we elaborate the numerical
algorithm for solving the GP equation in one space variable (the 1D, 
circularly-symmetric 2D, and spherically-symmetric 3D cases) 
and for calculating
the chemical potential, energy and rms sizes  employing both the real- and
imaginary-time propagation methods. 
In Sec. \ref{CN23D} we present the same  for
solving the GP equation in two and three space variables (the anisotropic  2D
and axially-symmetric and anisotropic 3D cases). In Sec. \ref{FOR} we present a
description of the Fortran programs, an explanation about how to use them, and
some sample outputs. In Sec. \ref{NUM} we present the numerical results for
chemical potential, rms size, value of the wave function at the center, for the
ground-state
problem using the imaginary-time propagation routines and compare our
finding with previous results for different trap symmetries in 1D, 2D, and 3D. 
We also present a study of non-stationary oscillation in some of these cases
using the real-time propagation routines when the nonlinear coefficient in the
GP equation with a stationary solution was suddenly reduced to half its value.
Finally, in Sec. \ref{SUM} we present a brief summary of our study.

\section{Nonlinear Gross-Pitaevskii Equation}
\label{gpe}

At zero temperature, the time-dependent Bose-Einstein condensate wave function
$\Psi \equiv \Psi({\bf r};\tau)$ at position ${\bf r}$ and time $\tau $ may be
described by the following  mean-field nonlinear GP equation \cite{review}
\begin{eqnarray}
\mbox{i}\hbar\frac{\partial \Psi({\bf r};\tau) }{\partial \tau}  =
\left[-\frac{\hbar^2\nabla
^2}{2m} + V({\bf r}) + gN\vert\Psi({\bf r};\tau) \vert^2
\right]\Psi({\bf r};\tau),
\label{eqn:gp0}
\end{eqnarray}
with $\mbox{i}=\sqrt{-1}$. Here $m$ is the mass of an atom and   $N$ the number
of atoms in the condensate, $g=4\pi\hbar^2 a/m $ the strength of inter-atomic
interaction, with $a$ the atomic scattering length. The normalization condition
of the wave function is $\int d{\bf r} \vert \Psi({\bf r};\tau)\vert^2 = 1$.

\subsection{Spherically-symmetric GP equation in 3D}

In this case  the trap potential is given by $V({\bf r}) =\frac{1}{2}m \omega ^2
\tilde r^2$, where $\omega$ is the angular frequency and $\tilde r$ the radial
distance. After a partial-wave projection the radial part $\psi$ of 
the wave function $\Psi$ can be written as $\Psi({\bf 
r};\tau) =\psi(\tilde
r,\tau)$. After a transformation of variables to  dimensionless quantities
defined by $r =\sqrt 2 \tilde r/l$, $t=\tau \omega$, $l\equiv \sqrt
{(\hbar/m\omega)} $ and $\phi(r;t) \equiv \varphi(r;t)/r =\psi(\tilde r,\tau)[
l^3/(2\sqrt 2)]^{1/2}$, the GP equation (\ref{eqn:gp0}) in this case becomes
\begin{eqnarray}
\left[-\frac{\partial^2}
{\partial r^2}+\frac{r^2}{4}+\aleph            
\left| \frac{\varphi(r;t)}{r}
\right| ^2 -
\mbox{i}\frac{\partial }{\partial t}\right] \varphi
(r;t)=0,
\label{sph}
\end{eqnarray}
where $\aleph 
=8\sqrt 2\pi  N a/l$. {The purpose of 
changing the wave function from $\psi$ to $\varphi=r\psi$ is a matter of 
taste and  it has certain advantages.
 First, this transformation removes the first derivative 
$\partial/\partial r$ from the differential equation (\ref{sph}) 
and thus results in a simpler equation  \cite{num26}. 
Secondly, at the origin $r=0$, $\psi$ is a constant, or 
$\partial \psi/\partial r=0$. But the new variable satisfies 
$\varphi(0,t)=0$. Hence, while solving the
differential equation (\ref{sph}),
we can implement the 
simple boundary condition 
that as $r \to 0 $ or $\infty$, $\varphi$ vanishes. 
The boundary condition for the 
differential 
equation in $\psi$ will be  a mixed one, e.g., the function $\psi$ 
should 
vanish at infinity  and its first space derivative should vanish at 
the origin. }     
The normalization condition for the
wave function is
\begin{align}
\label{n1}
4\pi \int_0^\infty dr \vert\varphi(r;t)\vert^2=1.
\end{align}

However, Eq. (\ref{sph}) is not the unique form of dimensionless GP equation in
this case. Other forms of dimensionless equations have been obtained and used by
different workers. For example, using  the transformations
$r =\tilde r/l$,
$t=\tau \omega$, $l\equiv \sqrt {(\hbar/m\omega)} $ and $\phi(r;t)
\equiv \varphi(r;t)/r =\psi(\tilde r,\tau)l^{3/2}$, the GP equation
(\ref{eqn:gp0}) becomes
\begin{align}
\left[-\frac{1}{2}\frac{\partial^2}
{\partial r^2}+\frac{1}{2}{r^2}+\aleph        
\left| \frac{\varphi(r;t)}{r}
\right| ^2 -
\mbox{i}\frac{\partial }{\partial t}\right] \varphi
(r;t)=0,
\label{sph2}
\end{align}
where $\aleph 
=4\pi  N a/l$ with normalization (\ref{n1}).
Finally, using
the transformations
$r =\tilde r/l$,
$t=\tau \omega/2$, $l\equiv \sqrt {(\hbar/m\omega)} $ and $\phi(r;t)
\equiv \varphi(r;t)/r =\psi(\tilde r,\tau)l^{3/2}$, the GP equation
(\ref{eqn:gp0}) becomes
\begin{align}
\left[-\frac{\partial^2}
{\partial r^2}+{r^2}+\aleph  
\left| \frac{\varphi(r;t)}{r}
\right| ^2 -
\mbox{i}\frac{\partial }{\partial t}\right] \varphi
(r;t)=0,
\label{sph3}
\end{align}
where $  \aleph 
=8\pi  N a/l$ with normalization (\ref{n1}). These three sets of
dimensionless GP equations have been {widely} used in the
literature and will be considered here. Equations (\ref{sph}), (\ref{sph2}), and
(\ref{sph3}) allow stationary solutions $\varphi(r;t)\equiv
\varphi(r)\exp(-\mbox{i}\mu t)$ where  $\mu$ is the chemical potential. The
boundary conditions for the solution of these equations are $\varphi(0,t)=0$ and
$\lim_{r\to \infty}\varphi(r,t)=0 $ \cite{koonin}.

\subsection{Anisotropic GP equation in 3D}

The three-dimensional trap potential is given by $V({\bf r})
=\frac{1}{2}m \omega^2(\nu^2\bar x^2+\kappa^2\bar
y^2+\lambda^2\bar
z^2)$,
where $\omega_x \equiv \nu \omega$, $\omega_y\equiv \omega\kappa$, and
$\omega_z\equiv \omega\lambda$ are the angular frequencies in the
$x$,
$y$ and $z$ directions, respectively, and ${\bf r}\equiv (\bar x,\bar
y,\bar z)$ is the radial vector. In terms of dimensionless variables
$x=\sqrt 2 \bar x/l, y=\sqrt 2 \bar y/l,z=\sqrt 2 \bar z/l, t=\tau
\omega, l=\sqrt{\hbar/(m\omega))}$, and $\varphi(x,y,z;t)=\sqrt{
l^3/(2\sqrt 2)}\Psi({\bf r};\tau)$, the GP equation (\ref{eqn:gp0})
becomes
\begin{align}
\left[
-\frac{\partial^2}{\partial  x^2}
-\frac{\partial^2}{\partial  y^2}
-\frac{\partial^2}{\partial  z^2}
+\frac{1}{4} \biggr(\nu^2x^2+\kappa^2 y^2+\lambda^2 z^2  \biggr)
+   \aleph 
\left\vert\varphi(x,y,z;t)\right\vert^2 - \mbox{i}
\frac{\partial}{\partial  t} \right]
\varphi(x,y,z;t)= 0,\label{ani}
\end{align}
with $\aleph 
=8\sqrt 2 \pi aN/l$ and normalization
\begin{align}\label{n3}
\int_{-\infty}^{\infty}dx
\int_{-\infty}^{\infty}dy
\int_{-\infty}^{\infty}dz
|\varphi(x,y,z;t)|^2 =1.
\end{align}

Similarly, using $x=\bar x/l, y=\bar y/l,z= \bar
z/l, t=\tau
\omega, l=\sqrt{\hbar/(m\omega)}$, and $\varphi(x,y,z;t)=\sqrt{
l^3}\Psi({\bf r};\tau)$, the GP equation (\ref{eqn:gp0}) becomes
\begin{eqnarray}
\left[
-\frac{1}{2}\frac{\partial^2}{\partial  x^2}
-\frac{1}{2}\frac{\partial^2}{\partial  y^2}
-\frac{1}{2}\frac{\partial^2}{\partial  z^2}
+\frac{1}{2} \biggr(\nu^2x^2+\kappa^2 y^2+\lambda^2 z^2  \biggr)
+\aleph  
\left\vert\varphi(x,y,z;t)\right\vert^2 - \mbox{i}
\frac{\partial}{\partial
t} \right]
\varphi(x,y,z;t)= 0,\label{ani2}
\end{eqnarray}
with $\aleph 
=4 \pi aN/l$ and normalization (\ref{n3}). Now 
with
scaling $t\to 2t$, Eq. (\ref{ani2}) can be rewritten as
\begin{eqnarray}
\left[
-\frac{\partial^2}{\partial  x^2}
-\frac{\partial^2}{\partial  y^2}
-\frac{\partial^2}{\partial  z^2}
+ \biggr(\nu^2x^2+\kappa^2 y^2+\lambda^2 z^2  \biggr)
+\aleph  
\left\vert\varphi(x,y,z;t)\right\vert^2 - \mbox{i}
\frac{\partial}{\partial
t} \right]
\varphi(x,y,z;t)= 0,\label{ani3}
\end{eqnarray}
with $\aleph  
=8\pi aN/l$.
The boundary conditions for solution are $\lim_{x\to \pm \infty}
\varphi(x,y,z;t)=0, \lim_{y\to \pm \infty}
\varphi(x,y,z;t)=0,\lim_{z\to \pm \infty}
\varphi(x,y,z;t)=0$  \cite{koonin}.

\subsection{Axially-symmetric GP equation in 3D}

In the special case of axial symmetry ($\nu = \kappa$) Eqs.
(\ref{ani}), (\ref{ani2}) and (\ref{ani3}) can be simplified considering
${\bf r}\equiv (\rho,z)$ where $\rho=\sqrt{x^2+y^2}$ is the radial
coordinate and $z$
is the axial coordinate. Then Eq. (\ref{ani}) becomes
\begin{align}
\left[
-\frac{\partial^2}{\partial  \rho^2}
-\frac{1}{\rho}\frac{\partial}{\partial  \rho}
-\frac{\partial^2}{\partial  z^2}
+\frac{1}{4} \biggr(\kappa^2 \rho^2+\lambda^2 z^2  \biggr)
+\aleph   
\left\vert\varphi(\rho,z;t)\right\vert^2 - \mbox{i}
\frac{\partial}{\partial  t} \right]
\varphi(\rho,z;t)= 0,\label{axi}
 \end{align}
with $\aleph  
=8\sqrt 2 \pi aN/l$ and normalization
$2\pi \int_{0}^{\infty}\rho d\rho
\int_{-\infty}^{\infty}dz
|\varphi(\rho,z;t)|^2 =1.$
Similarly, Eqs. (\ref{ani2}) and (\ref{ani3}) can be written as
\begin{eqnarray}
\left[
-\frac{1}{2}\frac{\partial^2}{\partial  \rho^2}
-\frac{1}{2\rho}\frac{\partial}{\partial  \rho}
-\frac{1}{2}\frac{\partial^2}{\partial  z^2}
+\frac{1}{2} \biggr(\kappa^2\rho^2+\lambda^2 z^2  \biggr)
+\aleph 
\left\vert\varphi(\rho,z;t)\right\vert^2 - \mbox{i}
\frac{\partial}{\partial
t} \right]
\varphi(\rho,z;t)= 0,\label{axi2}
\end{eqnarray}
with $\aleph  
=4 \pi aN/l$ and
\begin{eqnarray}
\left[
-\frac{\partial^2}{\partial  \rho^2}
-\frac{1}{\rho}\frac{\partial}{\partial  \rho}
-\frac{\partial^2}{\partial  z^2}
+ \biggr(\kappa^2 \rho^2+\lambda^2 z^2  \biggr)
+\aleph  
\left\vert\varphi(\rho,z;t)\right\vert^2 - \mbox{i}
\frac{\partial}{\partial
t} \right]
\varphi(\rho,z;t)= 0,\label{axi3}
\end{eqnarray}
with $\aleph   
=8\pi aN/l$. In this case $\varphi(\rho=0,z;t)$ 
is not
zero but a constant.
Convenient boundary conditions for solution
in this case are $\lim_{z\to \pm \infty}\varphi(\rho,z;t)=0, \lim_{\rho\to
\infty}\varphi(\rho,z;t)=0,$ and 
$\partial \varphi(\rho,z;t)/\partial \rho|_{\rho=0}=0$
\cite{num4}.

\subsection{One-dimensional GP equation}

In case of an elongated cigar-shaped trap, which is essentially an
axially-symmetric trap with strong transverse  confinement,  Eq.
(\ref{ani}) reduces to a quasi one-dimensional form. This is achieved by
assuming that the system remains confined to the ground state in the
transverse direction.  In this case the wave function of Eq.
(\ref{ani}) can be written as $\varphi(x,y,z;t) =\tilde \varphi(x;t)
\phi_0(y)
\phi_0(z)\exp[-i(\lambda+\kappa)t/2]$ with $\phi_0(y)=
[\kappa/(2\pi)]^{1/4} \exp(-\kappa y^2/4)$ and $\phi_0(z)=
[\lambda/(2\pi)]^{1/4} \exp(-\lambda z^2/4)$ the respective ground state
wave functions in $y$ and $z$ directions. Using this ansatz in Eq.
(\ref{ani}), multiplying by $\phi_0(y)\phi_0(z)$, integrating over
$y$ and $z$,  dropping the tilde over $\varphi$, and setting $\nu=1$  we
obtain
\begin{align}
\left[-\frac{\partial^2}
{\partial x^2}+\frac{x^2}{4}+\aleph 
\left| {\varphi(x;t)}
\right| ^2
-\mbox{i}\frac{\partial }{\partial t}
\right] \varphi (x;t)=0  ,
\label{1d}
\end{align}
with $\aleph 
= 2a N \sqrt {2
\lambda \kappa} /l$ and
 normalization
\begin{align}\label{n5}
\int_{-\infty}^\infty dx |\varphi(x;t)|^2 =1.
\end{align}
Instead if we employ
 $\varphi(x,y,z;t) =\tilde  \varphi(x;t) \phi_0(y)
\phi_0(z)\exp[-i(\lambda+\kappa)t/2]$ with $\phi_0(y)=
(\kappa/\pi)^{1/4} \exp(-\kappa y^2/2)$ and $\phi_0(z)=
(\lambda/\pi)^{1/4} \exp(-\lambda z^2/2)$
 in Eq.
(\ref{ani2}), in a similar fashion
 we obtain
\begin{align}
\left[-\frac{1}{2}\frac{\partial^2}
{\partial x^2}+\frac{x^2}{2}+\aleph  
\left| {\varphi(x;t)}
\right| ^2
-\mbox{i}\frac{\partial }{\partial t}
\right] \varphi (x;t)=0  ,
\label{1d2}
\end{align}
with $\aleph  
= 2a N \sqrt {
\lambda \kappa} /l$ and normalization (\ref{n5}). Now with scaling
$t\to 2t$ Eq. (\ref{1d2}) can be rewritten as
\begin{align}
\left[-\frac{\partial^2}
{\partial x^2}+{x^2}+\aleph  
\left| {\varphi(x;t)}
\right| ^2
-\mbox{i}\frac{\partial }{\partial t}
\right] \varphi (x;t)=0  ,
\label{1d3}
\end{align}
with $\aleph  
= 4a N \sqrt {
\lambda \kappa} /l$ and normalization (\ref{n5}).
For numerical solution we take $\lim_{x\to \pm \infty}\varphi(x,t)=0$.

\label{1D}

\subsection{Anisotropic GP equation in 2D}

\label{2.5}

In case of a disk-shaped trap, which is essentially an
anisotropic
trap in two dimensions
with strong axial binding Eq.
(\ref{ani}) reduces to a two-dimensional form. This is achieved by
assuming that the system remains confined to the ground state in the
axial direction.  In this case the wave function of Eq.
(\ref{ani}) can be written as $\varphi(x,y,z;t) = \tilde \varphi(x,y;t)
\phi_0(z)\exp[-i\lambda t/2]$ with  $\phi_0(z)=
[\lambda/(2\pi)]^{1/4} \exp(-\lambda z^2/4)$ the  ground state
wave function in  $z$ direction. Using this ansatz in Eq.
(\ref{ani}), multiplying by $\phi_0(z)$, integrating over
$z$, dropping the tilde over $\varphi$ and setting $\nu =1$  we obtain
\begin{align}
\left[-\frac{\partial^2}
{\partial x^2}-\frac{\partial^2}
{\partial y^2}
+\frac{x^2+\kappa y^2}{4}+\aleph 
\left| {\varphi(x,y;t)}
\right| ^2
-\mbox{i}\frac{\partial }{\partial t}
\right] \varphi (x,y;t)=0  ,
\label{2d}
\end{align}
now with $\aleph 
= 4a N \sqrt {2
\pi \lambda} /l$ and
 normalization
\begin{align}\label{n6}
\int_{-\infty}^\infty dx
\int_{-\infty}^\infty dy
|\varphi(x,y;t)|^2 =1.
\end{align}

Instead if we use in Eq. (\ref{ani2})
$\phi(x,y,z;t) =\tilde  \varphi(x,y;t)
\phi_0(z)\exp[-i\lambda t/2]$ with  $\phi_0(z)=
[\lambda/\pi]^{1/4} \exp(-\lambda z^2/2)$,  then
in a similar fashion  we obtain
\begin{align}
\left[-\frac{1}{2}\frac{\partial^2}
{\partial x^2}-\frac{1}{2}\frac{\partial^2}
{\partial y^2}
+\frac{x^2+\kappa y^2}{2}+\aleph  
\left| {\varphi(x,y;t)}
\right| ^2
-\mbox{i}\frac{\partial }{\partial t}
\right] \varphi (x,y;t)=0  ,
\label{2d2}
\end{align}
now with $\aleph  
= 2a N \sqrt {2
\pi \lambda} /l$ and normalization (\ref{n6}). Finally, with scaling
$t\to 2t$, Eq. (\ref{2d2}) can be written as
\begin{align}
\left[-\frac{\partial^2}
{\partial x^2}-\frac{\partial^2}
{\partial y^2}
+{x^2+\kappa y^2}+\aleph 
\left| {\varphi(x,y;t)}
\right| ^2
-\mbox{i}\frac{\partial }{\partial t}
\right] \varphi (x,y;t)=0  ,
\label{2d3}
\end{align}
with $\aleph 
= 4a N \sqrt {2
\pi \lambda} /l$ and normalization (\ref{n6}).
For numerical solution we take $\lim_{x\to \pm \infty}\varphi(x,y,t)=0$
and $\lim_{y\to \pm \infty}\varphi(x,y,t)=0$.

\subsection{Circularly-symmetric GP equation in 2D}

In the special case of circular symmetry the equations of Sec. \ref{2.5}
can be written in one-dimensional form. In this case $\kappa =1$, and
we introduce the radial variable ${\bf r}\equiv (x,y)$, and rewrite the
wave function as $\varphi(r)$. Then the GP equation (\ref{2d}) become
\begin{align}
\left[-\frac{\partial^2}
{\partial r^2}-\frac{1}{r}\frac{\partial}
{\partial r} +\frac{r^2}{4}+\aleph  
\left| {\varphi(r;t)}
\right| ^2 -
\mbox{i}\frac{\partial }{\partial t}\right] \varphi
(r;t)=0.
\label{cir}
\end{align}
 The normalization of  the
wave function is
$2\pi \int_0^\infty dr r\vert\varphi(r;t)\vert^2=1.$

In the circularly-symmetric case Eq. (\ref{2d2}) becomes
\begin{align}
\left[-\frac{1}{2}\frac{\partial^2}
{\partial r^2}-\frac{1}{2r}\frac{\partial}
{\partial r}+
\frac{1}{2}{r^2}+\aleph  
\left| {\varphi(r;t)}
\right| ^2 -
\mbox{i}\frac{\partial }{\partial t}\right] \varphi
(r;t)=0.
\label{cir2}
\end{align}
Finally, Eq. (\ref{2d3}) can be written as
\begin{align}
\left[-\frac{\partial^2}
{\partial r^2}-\frac{1}{r}\frac{\partial}
{\partial r}
+{r^2}+\aleph   
\left| {\varphi(r;t)}
\right| ^2 -
\mbox{i}\frac{\partial }{\partial t}\right] \varphi
(r;t)=0.
\label{cir3}
\end{align}
The convenient boundary condition in this case is $\lim_{r\to \infty }
\varphi
(r;t)=0$ and $d\varphi
(r;t)/dr|_{r=0}=0$ \cite{num4}.

In this section we have exhibited GP equations for different
trap symmetries. In the next section we illustrate the Crank-Nicolson
method for the GP equation in one space variable, which is then
extended to
other types of equations in Sec. \ref{CN23D}.

\section{Split-Step Crank-Nicolson Method for the GP
Equation in one Space Variable}
\label{CN1D}

\subsection{The GP Equation in the 1D and radially-symmetric 3D
cases}

To introduce the Crank-Nicolson Method \cite{koonin,ames,dtray}
for GP equation we consider
first the one-dimensional case of Sec. \ref{1D}. The
 nonlinear GP equation (\ref{1d})
in this
case can be expressed in the following form:
\begin{align}
\mbox{i}\frac{\partial }{\partial t} \varphi (x;t) & =
\left[-\frac{\partial^2}
{\partial x^2}+\frac{x^2}{4}+ \aleph 
\left| {\varphi(x;t)}
\right| ^2\right] \varphi (x;t)  , \notag \\
 & \equiv H  \varphi (x;t)
\label{e1d}
\end{align}
where the Hamiltonian $H$ contains the different linear and nonlinear terms
including the spatial derivative. (The spherically-symmetric GP equation in 3D
has a similar structure and can be treated similarly.) We solve this equation by
time iteration \cite{koonin,ames,dtray}. A given  trial input solution is
propagated in time over small time steps until a stable final  solution is
reached. The GP equation is discretized in space and time using the finite
difference scheme. This procedure results in a set of algebraic equations which
can be solved by time iteration using an input solution consistent with the
known boundary condition. In the present split-step method \cite{ames} this
iteration is conveniently done in few steps by breaking up the full Hamiltonian
into different derivative and non-derivative parts.

\subsubsection{Real-time propagation} \label{RT}

The time iteration is performed by splitting $H$ into two parts:
$H=H_1+H_2$, with
\begin{align}\label{h1}
H_1 & = \left[ \frac{x^2}{4}+\aleph  
\left|
{\varphi(x;t)}
\right| ^2 \right], \;\; \\ \label{h2}
H_2 & = -\frac {\partial ^2}{\partial x^2}.
\end{align}
{
Essentially we split Eq.~(\ref{e1d}) into
\begin{align}
\mbox{i}\frac{\partial }{\partial t} \varphi (x;t) & =
\left[ \frac{x^2}{4}+\aleph  
\left\vert {\varphi(x;t)}
\right\vert ^2\right] \varphi (x;t)  \equiv H_1  \varphi (x;t)
\label{e1d_1} \\
\mbox{i}\frac{\partial }{\partial t} \varphi (x;t) & =
-\frac{\partial^2}
{\partial x^2}\varphi (x;t)   \equiv H_2  \varphi (x;t)
\label{e1d_2} 
\end{align}
We first solve Eq.~(\ref{e1d_1}) with a initial value $\varphi (x;t_0)$ 
at $t = t_0$ to obtain an intermediate solution at $t = t_0 + \Delta$, where
$\Delta$ is the time step. 
Then this intermediate solution is used as initial value to solve 
Eq.~(\ref{e1d_2}) yielding the final solution at $t = t_0 + \Delta$ as 
$\varphi (x;t_0 + \Delta)$. This procedure is repeated $n$ times to get 
the final
solution at a given time $t_{\text{final}}=t_0+n\Delta $}. 

The time variable is discretized as $t_n=n\Delta$ where $\Delta$ is
the time step. The solution is advanced first over the time step
$\Delta$ at time $t_n$ by solving the GP equation (\ref{e1d}) with
$H=H_1$  to produce an intermediate solution $\varphi^{n+1/2}$ from
$\varphi^n$, where $\varphi^n$ is the discretized wave  function at
time $t_n$. As there is no derivative in $H_1$, this propagation is
performed essentially exactly for small $\Delta$ through the
operation
\begin{align}\label{al1}
\varphi^{n+1/2}
& = {\bigcirc}_{\mathrm{nd}}(H_1) \varphi^n \equiv  e^{-
\mbox{i}\Delta H_1}
\varphi^n,
\end{align}
where ${\bigcirc}_{\mathrm{nd}} (H_1)$ denotes time-evolution operation
with $H_1$ and the suffix `nd' denotes non-derivative. Next we perform
the time propagation corresponding to the operator $H_2$ numerically
by the semi-implicit Crank-Nicolson scheme (described below)
\cite{koonin}:
\begin{align}\label{gp2}
\frac{ \varphi^{n+1}- \varphi^{n+1/2}}{-\mbox{i}\Delta } =
\frac{1}{2}H_2(
\varphi^{n+1} + \varphi^{n+1/2}).
\end{align}
The formal solution to (\ref{gp2}) is
\begin{align}\label{gp3}
 \varphi^{n+1}= {\bigcirc}_{\mathrm{CN}}(H_2) \varphi^{n+1/2}
\equiv
\frac{1-\mbox{i}\Delta H_2/2  }{ 1+\mbox{i}\Delta
H_2/2 }
\varphi^{n+1/2},
\end{align}
which combined with Eq. (\ref{al1}) yields
\begin{align}\label{gp4}
 \varphi^{n+1}={\bigcirc}_{\mathrm{CN}}(H_2)   
{\bigcirc}_{\mathrm{nd}}(H_1)
\varphi^n,
\end{align}
where ${\bigcirc}_{\mathrm{CN}} $ denotes time-evolution operation with 
$H_2$
and   the suffix `CN' refers to the  Crank-Nicolson algorithm. Operation 
${\bigcirc}_{\mathrm{CN}} $ is used to propagate the intermediate 
solution  $ 
\varphi^{n+1/2} $ by time step $\Delta$  to generate the solution $ 
\varphi^{n+1}$ at the next time step $t_{n+1}=(n+1)\Delta$.

The advantage of the above split-step method with small time step $\Delta$ is
due to the following three factors \cite{ames,dtray}. First, all iterations
conserve normalization of the wave function. Second, the error involved in
splitting the Hamiltonian is proportional to $\Delta^2$ and can be neglected and
the method preserves the {symplectic} structure of the Hamiltonian formulation. 
Finally, as a major part of the Hamiltonian including the nonlinear term is
treated fairly accurately without mixing with the delicate Crank-Nicolson
propagation, the method can deal with an arbitrarily large nonlinear term and
lead to stable and accurate converged result.

Now we describe explicitly the semi-implicit
Crank-Nicolson algorithm.
The GP
equation is mapped onto  $N_x$ one-dimensional spatial grid points in
$x$.
Equation (\ref{e1d}) is discretized with $H=H_2$ of (\ref{h2}) by the
following Crank-Nicolson scheme \cite{koonin,ames,dtray}:
\begin{align}\label{kn1}
\frac{\mbox{i}(\varphi_{i}^{n+1}-
\varphi_{i}^{n+1/2})}{\Delta}=-\frac{1}{2h^2}\biggr[(\varphi^{n+1}_{i+1}-
2\varphi^{n+1}_i
+\varphi^{n+1}_{i-1})
+(\varphi^{n+1/2}_{i+1}-2\varphi_{i}^{n+1/2}
+\varphi^{n+1/2}_{i-1})\biggr],
\end{align}
where for the spherically-symmetric Eq. (\ref{sph})
$\varphi_i^n=\varphi(x_i;t_n)$ refers to  $x\equiv x_i=ih,$
$i=0,1,2,...,N_x$  and $h$ is the space 
step.
In the case of
the 
1D Eq. (\ref{1d}), we  choose  $x\equiv x_i=-N_xh/2+  ih,$
$i=0,1,2,...,N_x$.
 {(The choice $-$ even $N_x$ $-$ has the
advantage
of taking an equal number of space points on both sides of $x=0$ in the
1D  case setting the point $N_x/2$ at $x=0$.)} 
Equation (\ref{kn1}) is the
explicit form of the formal Eq. (\ref{gp2}). This scheme is
constructed by approximating $\partial /\partial t$ by a two-point
formula connecting the present ($n+1/2$) to future ($n+1$). The
spatial partial derivative
$\partial^2 /\partial x^2$ is
approximated  by a three-point formula averaged over the present and
the future time grid points.
This
procedure results in a series of tridiagonal sets of equations
(\ref{kn1}) in $\varphi^{n+1}_{i+1}$,  $\varphi^{n+1}_{i}$, and
$\varphi^{n+1}_{i-1}$ at time $t_{n+1}$, which are solved using the
proper boundary conditions.

The Crank-Nicolson scheme (\ref{kn1}) possesses certain properties
worth mentioning \cite{ames,dtray}. The error in this scheme is both
second order in space and time steps so that for small $\Delta$ and
$h$ the error is negligible. This scheme is also unconditionally
stable \cite{dtray}. The boundary condition at infinity is preserved for
small
values of $\Delta/h^2$\cite{dtray}.

The tridiagonal equations emerging from Eq. (\ref{kn1}) are written
explicitly as \cite{koonin}
\begin{align}\label{eq1}
A_i^-\varphi^{n+1}_{i-1}+A_i^0\varphi^{n+1}_{i}+
A_i^+\varphi^{n+1}_{i+1}= b_i,
\end{align}
where
\begin{align}
b_i=\frac{\mbox{i}\Delta}{2h^2}(\varphi^{n+1/2}_{i+1}-2\varphi_{i}^{n+1/2}
+\varphi^{n+1/2}_{i-1})+\varphi_i^{n+1/2},
\end{align}
and $A_i^-=A_i^+=  -\mbox{i}\Delta/(2h^2), A_i^0 = 1+\mbox{i}
\Delta/h^2$. All quantities in $b_i$ refer to time step $t_{n+1/2}$
and are considered known. The only unknowns in Eq. (\ref{eq1}) are the
wave forms $\varphi^{n+1}_{i\pm 1}$ and $\varphi^{n+1}_{i}$ at time
step $t_{n+1}$. To solve Eq. (\ref{eq1}), we assume the one-term
forward recursion relation
\begin{align}\label{eq2}
\varphi^{n+1}_{i+1}=\alpha_i\varphi^{n+1}_{i}+\beta_i,
\end{align}
where $\alpha_i$ and $\beta_i$ are coefficients to be determined.
Substituting Eq. (\ref{eq2}) in Eq. (\ref{eq1}) we obtain
\begin{align}\label{eq3}
A_i^-\varphi^{n+1}_{i-1}+A_i^0\varphi^{n+1}_{i}+
A_i^+(\alpha_i \varphi^{n+1}_{i}+\beta_i)= b_i,
\end{align}
which leads to the solution
\begin{align}\label{eq4}
\varphi_i^{n+1}=\gamma_i(A_i^- \varphi_{i-1}^{n+1}+A_i^+\beta_i-b_i),
\end{align}
with
\begin{align}\label{eq41}
\gamma_i=-1/(A_i^0+A_i^+\alpha_i).
\end{align}
From Eqs. (\ref{eq2}) and (\ref{eq4}) we obtain the following backward
recursion relations for the coefficients $\alpha_i$ and $\beta_i$
\begin{align}\label{eq5}
\alpha_{i-1}=\gamma_iA_i^-, \;\;\; \beta_{i-1}=
\gamma_i(A_i^+\beta_i-b_i).
\end{align}
We shall use the recursion relations (\ref{eq4}), (\ref{eq41}) and (\ref{eq5})
in a backward sweep of the lattice to determine $\alpha_i$ and $\beta_i$ for $i$
running from $N_x-2$ down to 0. The initial values chosen are $\alpha_{N_x-1}=0,
\beta_{N_x-1} = \varphi^{n+1}_{N_x}.$ This ensures the correct value of
$\varphi$ at the last lattice point. After determining the coefficients
$\alpha_i, \beta_i$ and $\gamma_i$, we can use the recursion relation
(\ref{eq2}) from $i=0$ to $N_x-1$ to determine the solution for the entire space
range using the starting value $\varphi^{n+1}_0$  (=0) known from the boundary
conditions. The value at the last lattice point is also taken to be known (= 0).
Thus we have determined the solution by using two sets of recursion relations
across the lattice each involving about $N_x$ operations.

In the numerical implementation of CN real-time propagation the
initial state at $t=0$ is usually chosen to be the analytically known
solution of the harmonic potential with zero nonlinearity: 
$
\aleph  
=0$. In the course of time iteration the nonlinearity is slowly
introduced until the desired final nonlinearity is attained. This
procedure will lead us to the final solution of the problem.

\subsubsection{Imaginary-time propagation}
\label{IT}

Although, real-time propagation as described above has many
advantages, in this approach one has to deal with complex variables
for a complex wave function for non-stationary states. For stationary
ground
state the wave function is essentially real- and the imaginary-time
propagation method dealing with real variables  seems to be
convenient. In this approach time $t$ is replaced by an imaginary
quantity $t=- \mbox{i}\bar t$ and Eq. (\ref{e1d}) now becomes
\begin{align}
-\frac{\partial }{\partial \bar  t} \varphi (x;\bar t) & =
\left[-\frac{\partial^2}
{\partial x^2}+\frac{x^2}{4}+\aleph  
\left|
{\varphi(x;\bar t)}
\right| ^2\right] \varphi (x;\bar t), \notag \\
& \equiv H  \varphi (x;\bar t)
\label{imt}
\end{align}
In this equation $\bar t$ is just  a mathematical parameter. 

{From Eq. (\ref{imt}) we see that an eigenstate 
$\varphi_i$ of eigenvalue $E_i$ 
of $H$ satisfying $H\varphi_i=E_i\varphi_i$ behaves under imaginary-time 
propagation as $\partial \varphi_i(\bar t) /\partial \bar t= -E_i 
\varphi_i(\bar t)$, so 
that  $\varphi_i(\bar t)= \exp(-E_i\bar t)\varphi_i( 0)$. Hence,  if we 
start 
with as arbitrary initial  $\varphi (x;\bar t)$ which can be taken as a 
linear combination of all eigenfunctions of $H$, then upon 
imaginary-time propagation all the eigenfunctions will decay 
exponentially with 
time. However, all the excited states with larger $E_i$ will decay 
exponentially faster compared to the ground state with the smallest 
eigenvalue. Consequently, after some time only the ground state 
survives. As all the states are decaying with time during 
imaginary-time propagation, we need to multiply the wave function by a 
number 
greater than unity to preserve its normalization so that the solution 
does not go to zero.}

The imaginary-time iteration is performed by splitting $H$ into two parts as
before: $H=H_1+H_2$, with $H_1$ and $H_2$ given by Eqs. (\ref{h1}) and
(\ref{h2}). It is realized that the entire analysis of Sec. \ref{RT} remains
valid provided we replace $\mbox{i}$ by 1 in Eq. (\ref{al1}) and by $-1$ in the
remaining equations. However, there appears one trouble.  The CN real-time
propagation preserves the normalization of the wave function, whereas the CN
imaginary-time propagation does not preserve the normalization. This problem can
be circumvented by restoring the normalization of the wave function after each
operation of Crank-Nicolson propagation. Once this is done the imaginary-time
propagation method for  stationary ground state problems yields very 
accurate result at
low computational cost.

Compared to the real-time propagation method, the imaginary-time
propagation method is very robust. The initial solution in the 
imaginary-time method 
could be any reasonable solution and not the analytically known solution
of a related problem as in the real-time method. { Also, the
full  nonlinearity can be added in a small number of time steps or even
in a single step and not in a large number of  steps as in the real-time 
method.} In the programs using the imaginary-time method  we include the 
 nonlinearity in a single step.  These two 
added features  together with
the use of real algorithm make the imaginary-time propagation method
very accurate with quick convergence for stationary ground states as we 
shall
see below.

\subsection{The GP Equation in the circularly-symmetric 2D
case}
\label{gp2dc}

The Crank-Nicolson discretization for real- and imaginary-time propagation for
the circularly-symmetric GP equation (\ref{cir}) is performed in a similar
fashion as for Eq. (\ref{e1d}), apart from the difference that here we also have
a first derivative in space variable in addition to the second derivative.
Another difference is that the wave function is not zero at $r=0$. In the case
of Eq. (\ref{e1d}) we took the boundary condition as $\varphi(x;t)=0$ at the
boundaries. For the circularly-symmetric case, the convenient boundary
conditions are $\lim_{r\to \infty}\varphi(r;t)=0$ and $d\varphi
(r;t)/dr|_{r=0}=0$.

We describe below the Crank-Nicolson discretization and the solution
algorithm in this case for the following equation
\begin{equation}
\left[-\frac{\partial^2}
{\partial r^2}-\frac{1}{r}\frac{\partial}
{\partial r}  -
\mbox{i}\frac{\partial }{\partial t}\right] \varphi
(r;t)=0,
\label{CNCIR}
\end{equation}
required for the solution of Eq. (\ref{cir}). The remaining procedure is
similar to that described in detail above in the one-dimensional case.

Equation (\ref{CNCIR}) is discretized by the
following Crank-Nicolson scheme as in Eq.
(\ref{kn1}) \cite{koonin,ames,dtray}:
\begin{align}\label{knx}
\frac{\mbox{i}(\varphi_{i}^{n+1}-
\varphi_{i}^{n+1/2})}{\Delta}=-\frac{1}{2h^2}\biggr[(\varphi^{n+1}_{i+1}-
2\varphi^{n+1}_i
+\varphi^{n+1}_{i-1})
+(\varphi^{n+1/2}_{i+1}-2\varphi_{i}^{n+1/2}
+\varphi^{n+1/2}_{i-1})\biggr]\nonumber \\ -\frac{1}{4r_ih}
\left[(\varphi^{n+1}_{i+1}-\varphi^{n+1}_{i-1})+(\varphi^{n+1/2}_{i+1}-
\varphi^{n+1/2}_{i-1})
\right],
\end{align}
where again
$\varphi_i^n=\varphi(r_i;t_n)$,   $r\equiv r_i=ih,$
$i=0,1,2,...,N_r$  and $h$ is the space step.
This scheme is
constructed by approximating $\partial /\partial x$ by a two-point
formula averaged over  present and  future time grid points.
The discretization of the first-order time and second-order space
derivatives is done as in Eq. (\ref{kn1}).
This
procedure results in the tridiagonal sets of equations
(\ref{knx}) in $\varphi^{n+1}_{i+1}$,  $\varphi^{n+1}_{i}$, and
$\varphi^{n+1}_{i-1}$ at time $t_{n+1}$, which are solved using the
proper boundary conditions.

The tridiagonal equations emerging from Eq. (\ref{knx}) are written
explicitly as
\begin{align}\label{ep1}
A_i^-\varphi^{n+1}_{i-1}+A_i^0\varphi^{n+1}_{i}+
A_i^+\varphi^{n+1}_{i+1}= b_i,
\end{align}
where
\begin{align}
b_i=\frac{\mbox{i}\Delta}{2h^2}(\varphi^{n+1/2}_{i+1}-2\varphi_{i}^{n+1/2}
+\varphi^{n+1/2}_{i-1})+\varphi_i^{n+1/2}+\frac{\mbox{i}\Delta}{4r_ih}
(\varphi^{n+1/2}_{i+1}-\varphi^{n+1/2}_{i-1}),
\end{align}
and $A_i^-= \mbox{i}\Delta[1/(4hr_i) -1/(2h^2)],
A_i^+= -\mbox{i}\Delta[1/(4hr_i) +1/(2h^2)],
A_i^0 =
1+\mbox{i}
\Delta/h^2$. All quantities in $b_i$ refer to time step $t_{n+1/2}$
and are considered known. The only unknowns in Eq. (\ref{ep1}) are the
wave forms $\varphi^{n+1}_{i\pm 1}$ and $\varphi^{n+1}_{i}$ at time
step $t_{n+1}$. To solve Eq. (\ref{ep1}), we assume the one-term
backward recursion relation
\begin{align}\label{ep2}
\varphi^{n+1}_{i-1}=\alpha_i\varphi^{n+1}_{i}+\beta_i,
\end{align}
where $\alpha_i$ and $\beta_i$ are coefficients to be determined.
Substituting Eq. (\ref{ep2}) in Eq. (\ref{ep1}) we obtain
\begin{align}\label{ep3}
A_i^+\varphi^{n+1}_{i+1}+A_i^0\varphi^{n+1}_{i}+
A_i^-(\alpha_i \varphi^{n+1}_{i}+\beta_i)= b_i,
\end{align}
which leads to the solution
\begin{align}\label{ep4}
\varphi_i^{n+1}=\gamma_i(A_i^+ \varphi_{i+1}^{n+1}+A_i^-\beta_i-b_i),
\end{align}
with
\begin{align}\label{ep41}
\gamma_i=-1/(A_i^0+A_i^-\alpha_i).
\end{align}
From Eqs. (\ref{ep2}) and (\ref{ep4}) we obtain the following
forward
recursion relations for the coefficients $\alpha_i$ and $\beta_i$
\begin{align}\label{ep5}
\alpha_{i+1}=\gamma_iA_i^+, \;\;\; \beta_{i+1}=
\gamma_i(A_i^-\beta_i-b_i).
\end{align}
We shall use the recursion relations (\ref{ep4}), (\ref{ep41}) and (\ref{ep5})
in a forward sweep of the lattice to determine $\alpha_i$ and $\beta_i$ for $i$
running from $1$  to $N_r-1$. The initial values chosen are $\alpha_{1}=1,
\beta_{1} =0.$ This ensures the correct value of the space derivative of
$\varphi(r;t)=0$ at $r=0$. After determining the coefficients $\alpha_i,
\beta_i$ and $\gamma_i$, we can use the recursion relation (\ref{ep2}) from
$i=N_r$ to $1$ to determine the solution for the entire space range using the
starting value $\varphi^{n+1}(N_r)$  (=0) known from the boundary condition.
Thus we have determined the solution by using two sets of recursion relations
across the lattice each involving about $N_r$ operations.

\subsection{Chemical potential}

\label{CH}

For stationary states the wave functions for the  1D case
have the trivial time
dependence $\varphi(x;t) \equiv \hat  \varphi(x) \exp(-\mbox{i}\mu
t)$,
where
$\mu $ is the chemical potential. Substituting this condition in Eq.
(\ref{1d}) we obtain
\begin{align} \label{mueq}
\left[-\frac{d^2}
{d x^2}+\frac{x^2}{4}+\aleph  
{\hat \varphi^2(x)}
\right] \hat \varphi (x)
=
  \mu \hat \varphi (x).
\end{align}
Assuming that the wave form is normalized to unity
$\int_{-\infty}^\infty \hat \varphi^2(x) dx =1$, the chemical
potential can
be calculated from the following expression obtained by multiplying
Eq. (\ref{mueq}) by $\hat \varphi (x)$ and integrating over all space
\begin{align}\label{muz}
\mu = \int_{-\infty}^\infty \left[\biggr(\frac{d\hat
\varphi(x)}{dx}
\biggr)^2 +\hat \varphi^2(x)
\left(
\frac{x^2}{4}+\aleph 
{\hat \varphi^2(x)}
\right) \right] dx,
\end{align}
where the second derivative has been simplified by an integration by
parts.

All the programs also calculate the many-body energy, which is of
interest. The analytical expression for energy is the
same as that of the chemical potential but with the nonlinear term
multiplied by 1/2, e.g., \cite{review}
\begin{align}\label{energy}
E = \int_{-\infty}^\infty \left[\biggr(\frac{d\hat
\varphi(x)}{dx}
\biggr)^2 +\hat \varphi^2(x)
\left(
\frac{x^2}{4}+\frac{\aleph} 
{2} {\hat \varphi^2(x)}
\right) \right] dx.
\end{align}
The programs will write the value of energy as output. However, we shall
not study or tabulate the results for energy and we shall not write the
explicit algebraic  expression of energy in the case of other trap
symmetries.

The GP equation (\ref{sph}) with spherically-symmetric potential  is
also an one-variable equation quite similar in structure to the
one-dimensional equation (\ref{1d}) considered above. Hence the entire
analysis of Secs. \ref{RT}, \ref{IT}, and \ref{CH} will be applicable
in this case
with $\varphi(x;t)$ replaced by $\varphi(r;t)/r$ in the
nonlinear term. Now if we consider stationary states of the form
$\varphi(r,t)\equiv \hat \varphi(r)\exp (-\mbox{i}\mu t/\hbar)$,
 the
expression for the chemical potential
becomes
\begin{align}\label{mueq2}
\mu =4\pi  \int_{0}^\infty \left[\biggr(\frac{d\hat \varphi(r)}{dr}
\biggr)^2 +\hat \varphi^2(r)
\left(
\frac{r^2}{4}+\aleph 
\frac{\hat \varphi^2(r)}{r^2}
\right) \right] dr.
\end{align}
We shall use Eqs. (\ref{muz}) and (\ref{mueq2}) for the calculation of
chemical potential from Eqs. (\ref{1d}) and (\ref{sph}). The energy will
be calculated from Eq. (\ref{energy}).
The expressions
for chemical potential in other cases can be written down in a
straight-forward fashion.

\section{Split-Step Crank-Nicolson method in two and three space
variables}
\label{CN23D}

\subsection{Anisotropic GP equation in 2D}

\label{anp}

In this case the GP equation (\ref{2d}) can be written as
\begin{eqnarray}
\mbox{i}
\frac{\partial}{\partial t}\varphi(x,y;t) &  =&
\biggr[
-\frac{\partial^2}{\partial  x^2}
-\frac{\partial^2}{\partial  y^2}
+ \frac{1}{4} \biggr( x^2 + \kappa^2 y^2   \biggr)
+\aleph  
|\varphi(x,y;t)|^2  \biggr]
\varphi(x,y;t) \nonumber \\ & \equiv & H \varphi(x,y;t)
,\label{a2}
\end{eqnarray}
with $\aleph  
= 4\sqrt {2 \pi\lambda} {Na}/{l}$.
The Hamiltonian $H$ can be conveniently broken
into three pieces $H=H_1+H_2+H_3$, where
\begin{align}
& H_1= \frac{1}{4} \biggr(x^2+\kappa^2 y^2  \biggr)
+\aleph  
|\varphi(x,y;t)|^2,  \\
& H_2=-\frac{\partial^2}{\partial x^2}, \;\;
H_3=-\frac{\partial^2}{\partial y^2}. \;\;
\end{align}
Now we adopt a policy quite similar to that elaborated in Sec.
\ref{RT} where the Hamiltonian was broken into two parts and where the
time propagation over time step $\Delta$ using the two parts were
carried out alternatively. The same procedure will be adopted in the
present case where we perform the time propagation using the pieces
$H_1$, $H_2$, and $H_3$ of the Hamiltonian successively in
three
independent time sub-steps $\Delta$ to complete a single time
evolution over time step $\Delta$ of the entire GP Hamiltonian $H$.
The time propagation over $H_1$ is performed as in Eq. (\ref{al1}) and
those over $H_2$ and $H_3$  as in Eqs. (\ref{gp3}) and
(\ref{kn1}).
The chemical potential in this case for the stationary state
$\varphi(x,y;t)\equiv \hat \varphi(x,y)\exp(-\mbox{i}\mu t)$
can be
written as
\begin{align}
\mu =
\int_{-\infty}^\infty dx \int_{-\infty}^\infty dy
 \left[ \biggr(\frac{{\partial}\hat \varphi}{
{\partial}x} 
\biggr)^2 +
\biggr(\frac{{\partial}\hat 
\varphi}{{\partial}y}
\biggr)^2
 + \hat \varphi^2 \left(
\frac{x^2+\kappa^2 y^2}{4}+\aleph  
{\hat \varphi^2} \right)
\right] ,
\end{align}
{
where we have performed integrations by parts 
to obtain this form for the chemical potential in terms of first 
derivatives only.}

\subsection{Axially-symmetric GP equation in 3D}

In this case the GP equation (\ref{axi}) can be written as
\begin{eqnarray}
\mbox{i}
\frac{\partial}{\partial t}\varphi(x,y;t) &  =&
\biggr[
-\frac{\partial^2}{\partial  \rho^2}
-\frac{1}{\rho}\frac{\partial}{\partial  \rho}
-\frac{\partial^2}{\partial  z^2}
+ \frac{1}{4} \biggr( \kappa^2\rho^2 + \lambda^2 z^2   \biggr)
+\aleph  
|\varphi(\rho,z;t)|^2  \biggr]
\varphi(\rho,z;t) \nonumber \\ & \equiv & H \varphi(\rho,z;t).
\label{ax2}
\end{eqnarray}
The Hamiltonian $H$ can be conveniently broken
into three pieces $H=H_1+H_2+H_3$, where
\begin{align}
& H_1= \frac{1}{4} \biggr(  \rho^2 + \lambda^2 z^2   \biggr)
+\aleph  
|\varphi(\rho,z;t)|^2,  \\
& H_2=-\frac{\partial^2}{\partial
\rho^2}-\frac{1}{\rho}\frac{\partial}{\partial  \rho}, \;\;
H_3=-\frac{\partial^2}{\partial z^2}. \;\;
\end{align}
Now we adopt a policy quite similar to that elaborated in Sec.
\ref{anp} where the Hamiltonian was broken into three parts and where
the
time propagation over time step $\Delta$ using the three parts were
carried out alternatively.
The time propagation over $H_1$ is performed as in Eq. (\ref{al1}) and
those over $H_2$ and $H_3$  as in Eqs. (\ref{knx}) and
(\ref{kn1}).
The chemical potential in this case for the stationary state
$\varphi(\rho,z;t)\equiv \hat \varphi(\rho,z)\exp(-\mbox{i}\mu t)$
can be
written as
\begin{align}
\mu = 2\pi
\int_{0}^\infty \rho d\rho \int_{-\infty}^\infty dz
 \left[ \biggr(\frac{{\partial}
\hat \varphi}{{\partial}\rho} 
\biggr)^2 +
\biggr(\frac{{\partial}\hat 
\varphi}{{\partial}z}
\biggr)^2
 + \hat \varphi^2 \left(
\frac{\kappa^2 \rho^2+\lambda^2 z^2}{4}+  \aleph 
{\hat 
\varphi^2}
\right)
\right] ,
\end{align}{
where we have again used integrations by parts to simplify the
final expression.}

\subsection{Anisotropic GP equation in 3D}

In this case the GP equation (\ref{ani}) can be written as
\begin{eqnarray}
\mbox{i}
\frac{\partial}{\partial t}\varphi(x,y,z;t) &  =&
\biggr[
-\frac{\partial^2}{\partial  x^2}
-\frac{\partial^2}{\partial  y^2}
-\frac{\partial^2}{\partial  z^2}
+ \frac{1}{4} \biggr(\nu^2 x^2 + \kappa^2 y^2 + \lambda^2 z^2  \biggr)
+\aleph   
|\varphi(x,y,z;t)|^2  \biggr]
\varphi(x,y,z;t) \nonumber \\ & \equiv & H \varphi(x,y,z;t)
,\label{a3}
\end{eqnarray}
with $\aleph 
= 8\sqrt 2 \pi {aN}/{l}$.
The Hamiltonian $H$ can be conveniently broken
into four pieces $H=H_1+H_2+H_3+H_4$, where
\begin{align}
& H_1= \frac{1}{4} \biggr(\nu^2 x^2 + \kappa^2 y^2 + \lambda^2 z^2
\biggr)
+\aleph  
|\varphi(x,y,z;t)|^2,  \\
& H_2=-\frac{\partial^2}{\partial x^2}, \;\;
H_3=-\frac{\partial^2}{\partial y^2}, \;\;
H_4=-\frac{\partial^2}{\partial z^2}.
\end{align}
Now we adopt a policy quite similar to that elaborated in Sec.
\ref{RT} where the Hamiltonian was broken into two parts and where the
time propagation over time step $\Delta$ using the two parts were
carried out alternatively. The same procedure will be adopted in the
present case where we perform the time propagation using the pieces
$H_1$, $H_2$, $H_3$ and $H_4$ of the Hamiltonian successively in four
independent time sub-steps $\Delta$ to complete a single time
evolution over time step $\Delta$ of the entire GP Hamiltonian $H$.
The time propagation over $H_1$ is performed as in Eq. (\ref{al1}) and
those over $H_2$, $H_3$ and $H_4$ as in Eqs. (\ref{gp3}) and (\ref{kn1}).
The chemical potential in this case for the stationary state
$ \varphi(x,y,z;t)\equiv \hat \varphi(x,y,z)\exp(-\mbox{i}\mu t)$
can
be
written as
\begin{align}
\mu =
\int_{-\infty}^\infty dx \int_{-\infty}^\infty dy \int_{-\infty}^\infty
dz \left[ \biggr(\frac{{\partial}
\hat \varphi}{{\partial}x} 
\biggr)^2 +
\biggr(\frac{{\partial}\hat 
\varphi}{{\partial}y}
\biggr)^2
+ \biggr(\frac{{\partial}\hat \varphi}
{{\partial}z} \biggr)^2 + 
\hat \varphi^2 \left(
\frac{\nu^2x^2+\kappa^2 y^2+\lambda^2 z^2}{4}+\aleph 
{\hat 
\varphi^2}
\right)
\right] ,
\end{align}{
where we have again used integrations by parts to simplify the
final expression.}

\section{Description of Numerical Programs}
\label{FOR}

\subsection{GP equation in one space variable}

In this subsection we describe six  Fortran  codes involving GP equation in one
space variable. These are programs for solving the 1D GP equation (program {\bf
imagtime1d.F}), the circularly-symmetric 2D GP equation (program {\bf
imagtimecir.F}) and the radially-symmetric 3D  GP equation  (program {\bf
imagtimesph.F}) using imaginary-time propagation. The similar routines using
real-time propagation are {\bf realtime1d.F}, {\bf realtimecir.F} and {\bf
realtimesph.F}, respectively. These real- and imaginary-time routines in 
one
space variable have similar structures. However, the wave function is real in
imaginary-time propagation, whereas it is complex in real-time propagation. To
accommodate this fact many variables in the real-time propagation routines are
complex.

The principal variables employed in the MAIN program are:  N = number of space
mesh points, NSTP = number of time iterations during which the nonlinearity is
introduced in real-time propagation, (in imaginary-time propagation the
nonlinearity is introduced in one step,) NPAS = number of subsequent time
iterations with fixed nonlinearity, NRUN =  number of final time iterations with
(a) fixed nonlinearity in imaginary-time propagation and (b) modified
nonlinearity to study dynamics in real-time propagation, X(I) = space mesh,
X2(I) = X(I)*X(I), V(I) = potential, CP(I) = wave function at space point X(I),
G, G0 = coefficient of nonlinear term, MU = chemical potential, EN = energy,
ZNORM = normalization of wave function, RMS = rms size or radius, DX = space
step, DT = time step, OPTION and XOP decide which equation to solve. Also
important is the subroutine INITIALIZE, where the space mesh X(I), potential  
V(I), and the initial wave function CP(I), are calculated.  (An advanced user
may need to change the variables V(I) and CP(I) so as to adopt the program to
solve  a different equation with different nonlinearity.) The functions and
variables not listed above are auxiliary variables, that the user should not
need to modify.

Now we describe the function of the subroutines, which  the user should not need
to change. The subroutine NORM calculates by Simpson's rule the normalization 
of the wave function and sets the normalization to unity, the preassigned value.
The real-time propagation preserves the normalization of the wave function and
hence the subroutine NORM is not used during time propagation. The subroutines
RAD and CHEM calculate the rms size (length or radius), and chemical 
potential (and energy) of
the system. The subroutines COEF and LU together implement the time propagation
with the spatial and temporal time derivative terms.  The subroutine NU performs
the time propagation with the nonlinear term and the potential.  In the
imaginary-time program the action of the subroutine LU does not preserve
normalization and hence each time the subroutine LU is called, the subroutine
NORM has to be called to set the normalization of the wave function back to
unity. (This is not necessary in the real-time programs which preserve the
norm.) The subroutine NONLIN calculates the nonlinear term. The subroutine DIFF
calculates the space derivative of the wave function by Richardson's
extrapolation formula needed for the computation of the chemical potential and
energy. The function SIMP does the numerical space integrations with the
Simpson's rule.

The programs implement the splitting method described in Sec.~\ref{RT} and
\ref{gp2dc} and calculate the wave function, chemical potential, size,
normalization, etc. The number of points in the one-dimensional space grid
represented by the integer variable N has to be chosen consistent with space
step DX such that the total space covered N$\times$DX is significantly larger
than the size of the condensate so that at the boundaries the wave 
function 
attains the asymptotic limits (e.g., the absolute value of the wave 
function  or its space derivative becomes 
less than 10$^{-10}$ or so 
for imaginary-time
propagation and less than 10$^{-7}$ for real-time propagation. Note 
that in the 1D and spherically-symmetric 3D problems we are using the 
asymptotic condition that the wave function is zero at both 
boundaries. In case of the circularly-symmetric 2D problem we use a mixed 
boundary condition, e.g.,  at origin the 
derivative of the wave function is zero and at infinity the wave 
function is zero.) In the
imaginary-time routine the total space covered should be about 1.5  times the
extension of the condensate. In the real-time routines, to obtain good 
precision the total space covered
should be at least 2  times the extension of the condensate. 
This is because the
imaginary-time routine is more precise and the solution attains its 
limiting asymptotic value at the
boundary very rapidly as the total space covered is increased. The real-time
routine is less accurate  and  one has to go to a larger distance before the
solution or its derivative drops to zero. A couple of runs with a 
sufficiently large N and large
DX are recommended to have an idea of the size/extension  of the system. 
A
smaller value of DX leads to a more accurate result provided an appropriate DT
is chosen.

The Crank-Nicolson method, described, for example, by Eq. (\ref{kn1}), is
unconditionally stable for all $\Delta/h^2$. Nevertheless, for a numerical
application to a specific problem one has to fix the time step DT  ($=\Delta$) and
space step DX ($=h$) for good convergence. Space and time steps are given in the
MAIN program,   as ``DATA DX/0.0025D0/, DT/0.00002D0/" with correlated DX and DT
values obtained by trial. (If the user wants to use other values of DX and DT, a
set of correlated values obtained by trial is given in Table \ref{table1}.)  The
total number of space points N in calculation has to be fixed in the line,
e.g., ``PARAMETER (N = 6000, N2 = N/2, NX = N-1)",  in the MAIN program, so
that the final wave function is within the space range and its value is
negligibly small at the boundaries.  The total numbers of time iterations are
fixed in the line, e.g., ``PARAMETER (NSTP = 500000, NPAS = 10000, NRUN =
20000)", in the MAIN program as described below in Sec. \ref{howto}.

The integer parameter OPTION should be set 1, 2 or 3 in the MAIN routine for
solution of equations of type  (\ref{sph3}),  (\ref{sph2}), or (\ref{sph}), [or
for solution of equations of type (\ref{1d3}), (\ref{1d2}), or (\ref{1d}),]
respectively. The difference between these three types of equations is in the
values of the coefficients in the first two terms. The nonlinear term calculated
in the subroutine   NONLIN is of the form $|\varphi(r,t)|^2$ with coefficient G.
If a different type of nonlinear term (with different functional dependence on
the wave function) is to be introduced, it should be done in the subroutine
NONLIN. Otherwise, this subroutine should not be changed. A different type of
nonlinear term is appropriate for a Fermi super-fluid \cite{ska1,ska2} or a
Tonks-Girardeau gas \cite{tg}.

\subsection{GP equation in two and three space variables}

In addition to the programs in one space variable we have six programs in two
and three space variables. The programs {\bf imagtime2d.F} and {\bf
realtime2d.F} apply to 2D Cartesian space using imaginary- and real-time
propagations, respectively. Similarly, {\bf imagtime3d.F} and {\bf realtime3d.F}
apply to 3D Cartesian space using imaginary- and real-time propagation,
respectively. Finally, programs {\bf imagtimeaxial.F} and {\bf realtimeaxial.F}
apply to an axially symmetric trap in 3D. The Fortran 90/95 versions of these
programs are somewhat more condensed and provide some advantage. Hence we also
include these programs as {\bf imagtime2d.f90},  {\bf realtime2d.f90}, 
{\bf
imagtime3d.f90},  {\bf realtime3d.f90}, {\bf imagtimeaxial.f90}, and 
{\bf
realtimeaxial.f90}. The output of the Fortran 90/95 programs are 
identical with
their corresponding Fortran 77 versions. All these programs are written using a
very similar logic used in the programs in one space variable. So most of the
considerations described earlier also applies to these cases. We describe the
principal differences below.

In case of total number of space points N, one now has the variables NX, NY, and
NZ for total number of space points in X, Y and Z directions in case of three
space variables. In case of two space variables the Z component is absent. These
variables
should be chosen equal to each other for a nearly  symmetric case. However, if
the problem is anisotropic in space, these variables can and should be chosen
differently. Similarly, the space variable X(I) is now replaced by space
variables X(I), Y(J), and Z(K) in three directions. Now there are space steps
DX, DY, and DZ in place of DX in one space variable. The variables DX, DY, and
DZ can now be chosen differently in three directions in case of an anisotropic
problem. However, they should  be chosen together with NX, NY, and NZ so that
the wave function lies entirely inside the chosen space and becomes negligibly
small at the boundaries. If the spatial extension of the wave function is much
smaller in one space direction than another, one should take a smaller space
step in the direction in which the spatial extension of the wave function is
small. The potential V and wave function CP      are now functions of 2 or 3
space variables and are represented by matrices in place of a column in the case
of one space variable. The variables AL (or KAP) and BL (or LAM) now define the
anisotropy of the trap and are used in the subroutine INITIALIZE to define the
trap. The subroutine LU is now replaced by LUX, LUY, and LUZ to implement the
solution in different space directions. In the imaginary-time propagation each
time the subroutines LUX, LUY, and LUZ  operate, one has to set the
normalization of the wave function to unity by calling the subroutine NORM.

\subsection{Instruction to use the programs}
\label{howto}

The programs, as supplied, solve the GP equations for a specific value of
nonlinearity and write the wave function, chemical potential, energy, rms size
or radius, wave function at the center, and nonlinearity, for  specific values
of space and time steps. The real- and imaginary-time programs for a 
specific
equation, for example, spherically-symmetric 3D equation, employ similar set of
parameters like space and time steps. The real-time programs use a larger value
of N, so that the discretization covers a larger region in space. In all cases
the supplied programs solve an equation of type (\ref{sph2}) with a factor of
1/2 in front of the space derivative, selected by setting the integer parameter
OPTION = 2 in the MAIN routine.  Other types of equations can be obtained by
setting OPTION = 1 for equations of type  (\ref{sph3}) or = 3 for equations of
type (\ref{sph}).

For solving a stationary ground state problem, the imaginary-time 
programs  are far more
accurate and should be used. The real-time programs should be used for studying
non-equilibrium problems often using  an initial wave function calculated by the
imaginary-time program. The non-equilibrium problems include the study of
soliton dynamics \cite{ska3}, expansion \cite{ska4}, collapse dynamics
\cite{ska5}, and other types of problems.

Each program is preset at a fixed nonlinearity G0 (= G), correlated DX-DT values
and NSTP, NPAS, and NRUN.  Smaller the steps DX and DT, more accurate will be
the result.  The correlated values of DX and DT on a data line should be found
by trial to obtain good convergence. 
Each supplied program produces result up-to a desired precision consistent
with the parameters employed $-$ G0, DX, DT, N, NSTP, NPAS, and NRUN. If the
nonlinearity G0 is increased, one might need to increase N to achieve similar
precision.
In many cases one may need an approximate solution (with lower
accuracy) involving less CPU time or one may need to solve the GP
equation for a different value of nonlinearity. 

(a) If G0 is reduced, just
change the card defining G0. However, if G0 is increased, changing the
value of G0 may not be enough.  For an increased G0, the wave function
extends to a larger region in space. One may need to increase the
``Number of space mesh points". For a new G0, just plot the output file
fort.3 for all programs except realtime3d.F, realtime3d.f90,
imagtime3d.F and imagtime3d.f90, where one should plot output files
fort.11, fort.12, and fort.13 and see that the wave function is fully
and adequately accommodated in the space domain. If not, one needs to
increase the number in input cards ``Number of space mesh points" until
the wave function is fully and adequately accommodated in the space
domain. 

(b)The CPU time involved can be reduced by sacrificing the precision.
This can be done by increasing the space step(s) and time step and
reducing the "Number of space mesh points". If the space step DX is
increased by a factor of f = 2, the number of space mesh points should
be reduced by the same factor. The time step DT should be increased by a
larger factor, more like f**2. The optimum increase in time step should
be determined by some experimentation. (A set of
correlated DX-DT
values is given in Tables \ref{table1} and \ref{table4} for the 
solution of corresponding equations.)
The CPU time is the largest in
the case of programs realtime3d.F, realtime3d.f90, imagtime3d.F and
imagtime3d.f90 and we give an example of changes below in these cases
to reduce the CPU time and precision. For example, for realtime3d.F, 
just change the 
lines  15, 16, 17, and 39.  The new
set of lines should be

      PARAMETER (NX=40, NXX = NX-1, NX2 = NX/2)

      PARAMETER (NY=32, NYY = NY-1, NY2 = NY/2)

      PARAMETER (NZ=24, NZZ = NZ-1, NZ2 = NZ/2)

      DATA DX /0.5D0/, DY /0.5D0/,  DZ /0.5D0/, DT/0.05D0/

For imagtime3d.F, just change the lines  15, 16, 17, and 37.  The new
set of lines should be

      PARAMETER (NX=24, NXX = NX-1, NX2 = NX/2)

      PARAMETER (NY=20, NYY = NY-1, NY2 = NY/2)

      PARAMETER (NZ=16, NZZ = NZ-1, NZ2 = NZ/2)

      DATA DX /0.5D0/, DY /0.5D0/,  DZ /0.5D0/, DT/0.04D0/

Please verify, by running the corresponding programs,  that the new 
results so obtained are quite similar to the 
existing results.

The integer parameter NSTP refers to the number of time iterations during which
the nonlinear term is slowly introduced in the real-time propagation. This
number should be large (typically more than 100,000 for small nonlinearity, for
larger nonlinearity could be 1000,000 ) for good convergence; this means that
the nonlinearity should be introduced in small amounts over a large number of
time iterations. In real-time propagation, NPAS refers to certain number of time
iterations with the constant nonlinear term already introduced in NSTP  and
should be small (typically 1000). NRUN refers to time iterations with a 
modified
nonlinearity so as to generate a non-equilibrium dynamics. In the imaginary-time
propagation the parameters NPAS and NRUN refer to certain number of time
iterations with the constant nonlinear term already introduced in one step and
should be large (typically NPAS = 200,000 or more) for good convergence.

\subsection{Output files}

The programs write via statements WRITE(1,*), WRITE(2,*) WRITE(3,*)  in Files 1,
2, and 3, respectively, the initial stationary wave function  and that after
NSTP, and NPAS time iterations for real-time propagation and after NPAS and NRUN
time iterations for imaginary-time propagation. File 3  gives the final
stationary wave function of the calculation. However, in the case of the
anisotropic 3D programs realtime3d.F and imagtime3d.F, sections of the wave
function as plotted in Fig. \ref{fig4} (b) are written in Files 1, 2, and 3
before  NSTP (realtime3d.F) and after NPAS  (imagtime3d.F) iterations, 
and in Files 11,
12, and 13 after NPAS (realtime3d.F) and NRUN  (imagtime3d.F) iterations.

In the real-time program a non-stationary oscillation is initiated by suddenly
modifying the nonlinearity from G to G/2 after NPAS time iterations. During NRUN
time iterations the non-stationary dynamics is studied. The real-time programs
write on File 8  the running time and  rms size during non-stationary
oscillation.

In addition, these programs write in File 7 the values of nonlinearity G, space
steps DX, DY, DZ,  time step DT, number of space mesh points N,  number of time
iterations NSTP, NPAS, and NRUN together with the values of normalization of the
wave function, chemical potential, energy, rms size, value of the wave function
at center, nonlinearity coefficient G.

Below we provide some sample output listed on File 7 from the different programs
using OPTION = 2. File 7 (fort.7) represents the comprehensive result in each
case.

(1)  Program {\bf imagtime1d.F}

\begin{verbatim}

  OPTION =   2

# Space Stp N =     8000
# Time Stp :  NPAS =    200000, NRUN =     20000
  Nonlinearity G =      62.74200000
  Space Step DX =     0.002500, Time Step DT =     0.000020

                   ----------------------------------------------------
                    Norm       Chem        Ener       <x>       Psi(0)
                   ----------------------------------------------------
Initial :          1.0000    0.500000    0.500000    0.70711    0.75113
After NPAS iter.:  0.9996   10.369462    6.256976    2.04957    0.40606
After NRUN iter.:  0.9996   10.369462    6.256976    2.04957    0.40606
                   ----------------------------------------------------

\end{verbatim}
(2)  Program {\bf imagtimesph.F}

\begin{verbatim}

  OPTION =   2

# Space Stp N =     3000
# Time Stp : , NPAS =    200000, NRUN =     20000
  Nonlinearity G =     125.48400000
  Space Step DX =     0.002500, Time Step DT =     0.000020

                   ----------------------------------------------------
                    Norm       Chem        Ener       <r>       Psi(0)
                   ----------------------------------------------------
Initial :          1.0000    1.500000    1.500000    1.22474    0.42378
After NPAS iter.:  0.9998    4.014113    3.070781    1.88214    0.17382
After NRUN iter.:  0.9998    4.014113    3.070781    1.88214    0.17382
                   ----------------------------------------------------
\end{verbatim}

(3)  Program {\bf imagtimecir.F}
\begin{verbatim}

  OPTION =   2

# Space Stp N =     2000
# Time Stp : , NPAS =    200000, NRUN =    200000
  Nonlinearity G =      -2.50970000
  Space Step DX =     0.002500, Time Step DT =     0.000020

                   ----------------------------------------------------
                    Norm       Chem        Ener       <r>       Psi(0)
                   ----------------------------------------------------
Initial :          1.0000    1.000000    1.000000    1.00000    0.56419
After NPAS iter.:  1.0000    0.499772    0.770107    0.87758    0.67535
After NRUN iter.:  1.0000    0.499772    0.770107    0.87758    0.67535
                   ----------------------------------------------------

\end{verbatim}

(4)  Program {\bf imagtime2d.F and imagtime2d.f90}
\begin{verbatim}


  OPTION =   2
  Anisotropy AL =     2.000000

# Space Stp NX =      800, NY =      800
# Time Stp : NPAS =     30000, NRUN =      5000
  Nonlinearity G =      12.54840000
  Space Step DX =   0.020000, DY =   0.020000
  Time Step  DT =   0.000100

                   -----------------------------------------------------
                    Norm       Chem        Ener       <r>       Psi(0,0)
                   -----------------------------------------------------
Initial :          1.0000    1.500000    1.500000    0.86603    0.67094
After NPAS iter.:  0.9999    3.254878    2.490493    1.17972    0.46325
After NRUN iter.:  0.9999    3.254878    2.490493    1.17972    0.46325
                   -----------------------------------------------------

\end{verbatim}

(5)  Program {\bf imagtimeaxial.F and imagtimeaxial.f90}
\begin{verbatim}
  OPTION =   2
  Anisotropy KAP =     1.000000, LAM =     4.000000

# Space Stp NX =      500, NY =      500
# Time Stp : NPAS =     100000, NRUN =     20000
  Nonlinearity G =      18.81000000
  Space Step DX =   0.020000, DY =   0.020000
  Time Step  DT =   0.000040

            ------------------------------------------------------------------
             Norm    Chem      Energy    <rho>      <z>       <r>     psi(0)
            ------------------------------------------------------------------
Initial   : 1.000   3.00000   3.00000   1.00000   0.35355   1.06066   0.59883
NPAS iter : 1.000   4.36113   3.78228   1.32490   0.38049   1.37846   0.38129
NRUN iter : 1.000   4.36113   3.78228   1.32490   0.38049   1.37846   0.38129
            ------------------------------------------------------------------

\end{verbatim}

(6) Program {\bf imagtime3d.F and imagtime3d.f90}
\begin{verbatim}
 OPTION =   2
  Anisotropy AL =     1.414214, BL = 2.000000

# Space Stp NX =      240, NY =      200, NZ =      160
# Time Stp : NPAS =      5000, NRUN =       500
  Nonlinearity G =      44.90700000
  Space Step DX =   0.050000, DY =   0.050000, DZ =   0.050000
  Time Step  DT =   0.000400

                  ------------------------------------------------------
                    Norm      Chem        Ener        <r>     Psi(0,0,0)
                  ------------------------------------------------------
Initial         :  1.0000      2.2071      2.2071     1.0505     0.5496
After NPAS iter.:  0.9997      4.3446      3.4862     1.4583     0.2888
After NRUN iter.:  0.9997      4.3446      3.4862     1.4583     0.2888
                  ------------------------------------------------------


\end{verbatim}

(7)  Program {\bf realtime1d.F}
\begin{verbatim}
  OPTION =   2

# Space Stp N =     5000
# Time Stp : NSTP  1000000 , NPAS =     1000, NRUN =     40000
  Nonlinearity G =      62.74200000
  Space Step DX =     0.010000, Time Step DT =     0.000100

                   ----------------------------------------------------
                    Norm       Chem        Ener       <x>       Psi(0)
                   ----------------------------------------------------
Initial :          1.0000       0.500       0.500      0.707      0.751
After NSTP iter.:  1.0000      10.368       6.257      2.050      0.406
After NPAS iter.:  1.0000      10.375       6.257      2.047      0.406
                   ----------------------------------------------------



\end{verbatim}

(8)  Program {\bf realtimesph.F}
\begin{verbatim}
  OPTION =   2

# Space Stp N =     2000
# Time Stp : NSTP =   1000000, NPAS =      1000, NRUN =     40000
  Nonlinearity G =     125.48400000
  Space Step DX =     0.010000, Time Step DT =     0.000100

                   ----------------------------------------------------
                    Norm       Chem        Ener       <r>       Psi(0)
                   ----------------------------------------------------
Initial :          1.0000       1.500       1.500      1.225      0.424
After NSTP iter.:  1.0000       4.015       3.071      1.881      0.174
After NPAS iter.:  1.0000       4.011       3.071      1.884      0.174
                   ----------------------------------------------------

\end{verbatim}

(9)  Program {\bf realtimecir.F}
\begin{verbatim}
  OPTION =   2

# Space Stp N =     2000
# Time Stp : NSTP =   1000000, NPAS =      1000, NRUN =     40000
  Nonlinearity G =      12.54840000
  Space Step DX =     0.010000, Time Step DT =     0.000100

                   ----------------------------------------------------
                    Norm       Chem        Ener       <r>       Psi(0)
                   ----------------------------------------------------
Initial :          1.0000       1.000       1.000      1.000      0.564
After NSTP iter.:  1.0000       2.255       1.708      1.308      0.391
After NPAS iter.:  1.0000       2.255       1.708      1.308      0.391
                   ----------------------------------------------------
\end{verbatim}

(10)  Program {\bf realtime2d.F and realtime2d.f90}
\begin{verbatim}

  OPTION =   2
  Anisotropy AL =     1.000000

# Space Stp NX =      200, NY =      200
# Time Stp : NSTP =    100000, NPAS =      1000, NRUN =      5000
  Nonlinearity G =      12.54840000
  Space Step DX =   0.100000, DY =   0.100000
  Time Step  DT =   0.001000

                   -----------------------------------------------------
                    Norm       Chem        Ener       <r>       Psi(0,0)
                   -----------------------------------------------------
Initial :           1.000       1.000       1.000      1.000      0.564
After NSTP iter.:   1.000       2.256       1.708      1.307      0.392
After NPAS iter.:   1.000       2.257       1.708      1.305      0.392
                   -----------------------------------------------------





\end{verbatim}

(11)  Program {\bf realtimeaxial.F and realtimeaxial.f90}
\begin{verbatim}

  OPTION =   2
  Anisotropy KAP =     1.000000, LAM =     4.000000

# Space Stp NX =      130, NY =      130
# Time Stp : NSTP =    100000, NPAS =      1000, NRUN =     20000
  Nonlinearity G =      18.81000000
  Space Step DX =   0.100000, DY =   0.100000
  Time Step  DT =   0.001000

            -------------------------------------------------------
              Norm    Chem     Energy    <rho>     <z>       psi(0)
            -------------------------------------------------------
Initial :   1.000     3.000     3.000     1.000     0.354     0.587
NSTP iter : 0.999     4.362     3.782     1.323     0.381     0.376
NPAS iter : 1.000     4.362     3.782     1.327     0.379     0.376
            -------------------------------------------------------


\end{verbatim}

(12)  Program {\bf realtime3d.F and realtime3d.f90}
\begin{verbatim}
  OPTION =   2
  Anisotropy AL =     1.414214, BL =     2.000000

# Space Stp NX =      200, NY =      160, NZ =      120
# Time Stp : NSTP =     60000, NPAS =      1000, NRUN =      4000
  Nonlinearity G =      22.45400000
  Space Step DX =   0.100000, DY =   0.100000, DZ =   0.100000
  Time Step  DT =   0.002000

                  ------------------------------------------------------
                    Norm      Chem        Ener        <r>     Psi(0,0,0)
                  ------------------------------------------------------
Initial :           1.000      2.207      2.207     1.051     0.550
After NSTP iter.:   1.000      3.572      2.992     1.321     0.347
After NPAS iter.:   1.000      3.572      2.992     1.320     0.347
                  ------------------------------------------------------


\end{verbatim}

\label{5.3}

\section{Numerical Results}
\label{NUM}
\subsection{Stationary Problem}

\begin{table}[!ht]
\begin{center}
\caption{Convergence of result in the   1D (X) and
radially-symmetric  3D ($r$) cases for  nonlinearities $\aleph 
=627.42$ and
627.4, calculated
using the imaginary-time propagation programs imagtime1d.F  [Eq.
(\ref{1d2}), OPTION 2]
and imagtimesph.F [Eq. (\ref{sph2}), OPTION 2]
respectively, for various space step DX and time step DT.
}
\label{table1}
\begin{tabular}{|r|r|r|r|r|r|}
\hline
$\aleph $  
& DX  &  DT &
$\varphi(0)/\phi(0)$ &  $x_{\mathrm{rms}}/r_{\mathrm{rms}}$ &
{$\mu$ } \\
\hline
  627.42(X) & 0.08  & 0.005   & 0.276647 &   4.384825 & 48.024062  \\
  627.42(X) & 0.04  & 0.001  &0.276648 &4.384744 & 48.024389  \\
  627.42(X) & 0.02  &0.0005   &0.276649&     4.384734  & 48.024429  \\
  627.42(X) & 0.01 &  0.0001  &0.276649&    4.384726 &  48.024462  \\
  627.42(X) & 0.005  & 0.00005  &0.276649   &     4.384725 &
48.024466  \\
  627.42(X) & 0.0025  & 0.00002  &0.276649   &     4.384724 &
48.024468  \\
  627.42(X) & 0.001  &   0.00001  & 0.276649   &     4.384724 &
48.024468
\\
\hline
  627.4($r$) &0.08   &   0.005   &   0.106655 &      2.506348   &
7.247479
\\
  627.4($r$) &0.04   &    0.001  & 0.106679    &   2.505886    &
7.248206
\\
  627.4($r$) & 0.02  & 0.0005  & 0.106684&       2.505833  &  7.248292
\\
  627.4($r$) & 0.01& 0.0001  &0.106686&        2.505785     & 7.248365
\\
  627.4($r$) & 0.005  &  0.00005  &0.106686 &       2.505780  & 7.248374
\\
  627.4($r$) & 0.0025  &0.00002  &0.106686 &       2.505776    &
7.248380  \\
  627.4($r$) & 0.001  & 0.00001  & 0.106686   &    2.505776    &
7.248380  \\
\hline
\end{tabular}
\end{center}
\end{table}

\begin{figure}[tbp] \begin{center}
{\includegraphics[width=.8\linewidth]{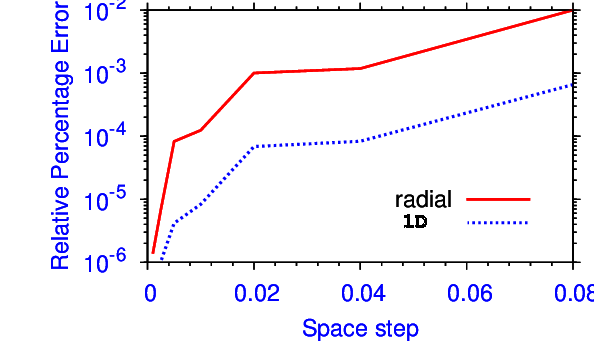}}
\end{center}
\caption{(Color online) Relative percentage error
as a function of
space step
for one-dimensional (1D)  and spherically-symmetric  3D (radial)
models with nonlinearity $\aleph 
=627.42$ and 627.4, respectively,
calculated using the imaginary-time propagation programs imagtime1d.F
[Eq. (\ref{1d2}), OPTION 2]
and imagtimesph.F [Eq. (\ref{sph2}), OPTION 2]
with the
data of
Table \ref{table1}.
}
\label{fg1}
\end{figure}

\begin{table}[!ht]
\begin{center}
\caption{The chemical potential $\mu$, rms radius {$r_{\mathrm{rms}}$}, and the wave function
$\phi(0)$ at the center  for various
nonlinearities in the radially symmetric 3D case calculated using the
program imagtimesph.F [Eq. (\ref{sph2}), OPTION 2]. Table completed with space step
DR $
\le 0.0025$
and DT $=0.00002$. }
\label{table2}
\begin{tabular}{|r|r|r|r|r|r|r|}
\hline
{$ \aleph $
} &   {$\phi(0)$} &
$\phi(0)$ \cite{Bao_Tang}&
{$r_{\mathrm{rms}}$}      &
$r_{\mathrm{rms}}$ \cite{Bao_Tang}&
{$\mu$}  &{$\mu$ \cite{Tiwari_Shukla,Bao_Tang}} \\
\hline
-3.1371 & 0.48792(1) &  0.4881 & 1.51213(1)& 1.1521& 1.265184(2) &
$1.2652$ \\
       0 &  0.42378 & 0.4238& 1.22474& 1.2248 & 1.500000 &1.5000
\\
    3.1371 &0.38425(1)& 0.3843 &1.27857(1)   &1.2785 & 1.677451(1)  &
1.6774\\
   12.5484 & 0.31800(1)  & 0.3180 &  1.39211(1)  & 1.3921 &
2.065018(1) &
2.0650 \\
  31.371 & 0.25810(1)&  0.2581  & 1.53561(1)& 1.5356  &  2.586116(1) &
2.5861 \\
125.484 &  0.17382(1) &  0.1738 &  1.88215(1) &1.8821 &  4.014113(2)
&4.0141 \\
  627.4 & 0.10669(1) &0.1066 & 2.50578(1) &2.5057 & 7.248380(3)  &
7.2484
\\
    3137.1 &0.06559(1) &0.0655 & 3.41450(1) &3.4145 &  13.553403(4)  &
13.553
\\
\hline
\end{tabular}
\end{center}
\end{table}

In this subsection we present results for the stationary ground 
state problem
calculated with the imaginary-time programs.
All numerical results presented in this paper are for the
nonlinear equations with a factor of 1/2 in front of the gradient term
obtained by choosing OPTION = 2 in the MAIN.
First we consider the numerical result for the simplest cases $-$ the
1D and the radially-symmetric 3D problems by
imaginary-time propagation with nonlinearities $\aleph 
=$ 627.42 
and
627.4, respectively. The calculations were performed with different
space and time steps DX and DT, respectively. (As DX is reduced, DT
should be reduced also to have good convergence. The correlated DX-DT
values were obtained by trial to achieve good convergence.)
In each case a
sufficiently large number of space points N is to be taken, so that the
space domain of integration covers the extension of the wave function adequately.
We exhibit in Table \ref{table1}, the results for the wave function at
center, rms
size, and chemical potential in these two cases for a fixed
nonlinearity for different DX and DT.
We find that convergence is achieved up to six significant digits after the
decimal point
with space step DX = 0.0025 and time step DT = 0.00002.
In Fig. \ref{fg1} we plot the relative percentage error in chemical
potential $\mu$ for various space steps. The percentage error rapidly
reduces as space step is reduced.

In Table \ref{table2} we exhibit the chemical potential, rms radius, and
the wave function at the center for various nonlinearities  in the
spherically symmetric 3D case using space step
DX = 0.0025 and time step DT = 0.00002 in the imaginary-time
propagation program. From the
table we find that the results are in agreement with those calculated
in Refs. \cite{Tiwari_Shukla,Bao_Tang}.

\begin{table}[!ht]
\begin{center}
\caption{The chemical potential $\mu$,
rms size {$x_{\mathrm{rms}}$}, and the wave function
$\varphi(0)$ at the center  for various
nonlinearities
 in the  1D case calculated using the program imagtime1d.F
[Eq. (\ref{1d2}), OPTION 2]. Table
completed with DX $\le 0.0025$ and
DT $=0.00002$.
}
\label{table3}
\begin{tabular}{|r|r|r|r|r|r|r|}
\hline
{$\aleph$} 
&   {$\varphi(0)$} &
$\varphi(0)$ \cite{Bao_Tang}&
{$x_{\mathrm{rms}}$}      &
$x_{\mathrm{rms}}$ \cite{Bao_Tang}&
{$\mu$}  &{$\mu$ \cite{Bao_Tang}} \\
\hline
  -2.5097 & 0.91317(1) &  0.9132 & 0.51334(1)& 0.5133& $-0.80623(3)$ &
$-0.8061$ \\
   0 &  0.75112 & 0.7511& 0.70711& 0.7071 & 0.500000 &0.5000    \\
    3.1371 &0.64596(1)& 0.6459 &0.89602(1)&0.8960 &$1.526593(3)$  &
1.5265\\
   12.5484 & 0.52975(1)& 0.5297 & 1.24549(1)& 1.2454 &   3.596560(2) &
3.5965 \\
  31.371 & 0.45567(1)&0.4556  & 1.64170(1)&1.6416 &  $6.552682(2)$ &
6.5526 \\
62.742&  0.40606(1)& 0.4060 &  2.04957(1) & 2.0495 & $10.369462(2)$ &
10.369
\\
156.855 &  0.34856(1) &  0.3485 &  2.76794(1) &2.7679 &$19.070457(2) $
&19.0704 \\
   313.71 & 0.31053(1) & 0.3105 &3.48237(1) &3.4823 & 30.259178(3) &
30.259
\\
  627.42 & 0.27665(1) &0.2766 & 4.38472(1) &4.3847 & 48.024468(3)  &
48.024
\\
     1254.8 & 0.24647(1) &0.2464 & 5.52282(1) &5.5228 & 76.226427(3)&
76.226
\\
\hline
\end{tabular}
\end{center}
\end{table}

In Table \ref{table3} we exhibit the chemical potential, rms size, and
the wave function at the center for various nonlinearities  in the
 1D  case using space step
DX = 0.0025 and time step DT = 0.00002. From the
table we find that the results are in agreement with those calculated
in Ref. \cite{Bao_Tang}.

\begin{figure}[tbp] \begin{center}
{\includegraphics[width=.49\linewidth]{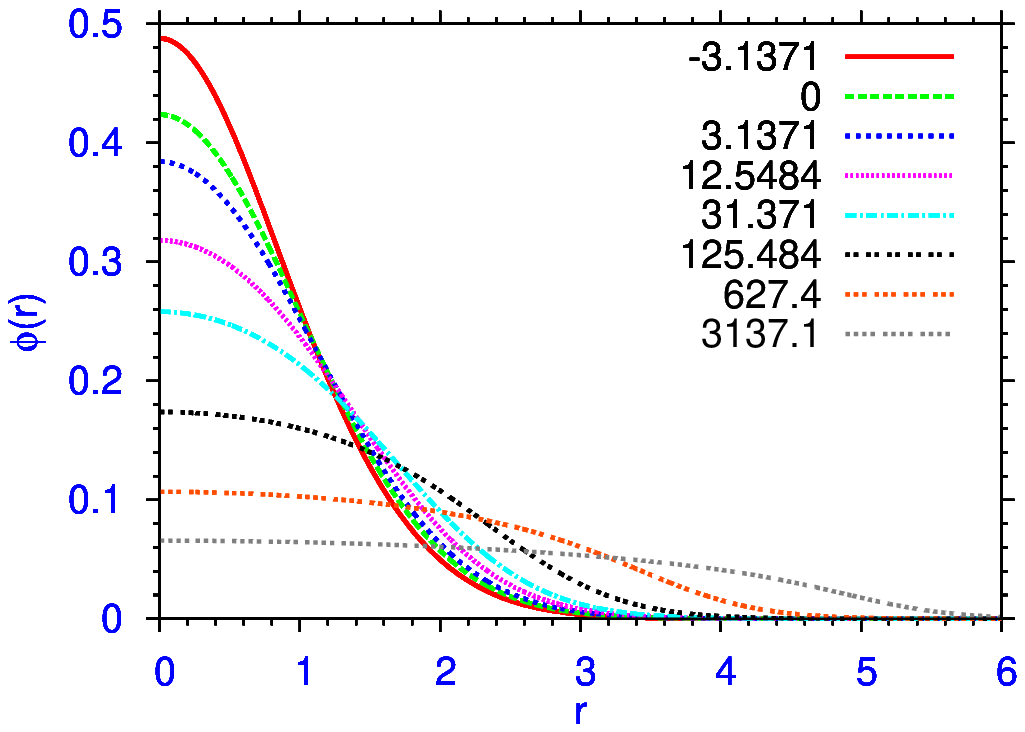}}
{\includegraphics[width=.49\linewidth]{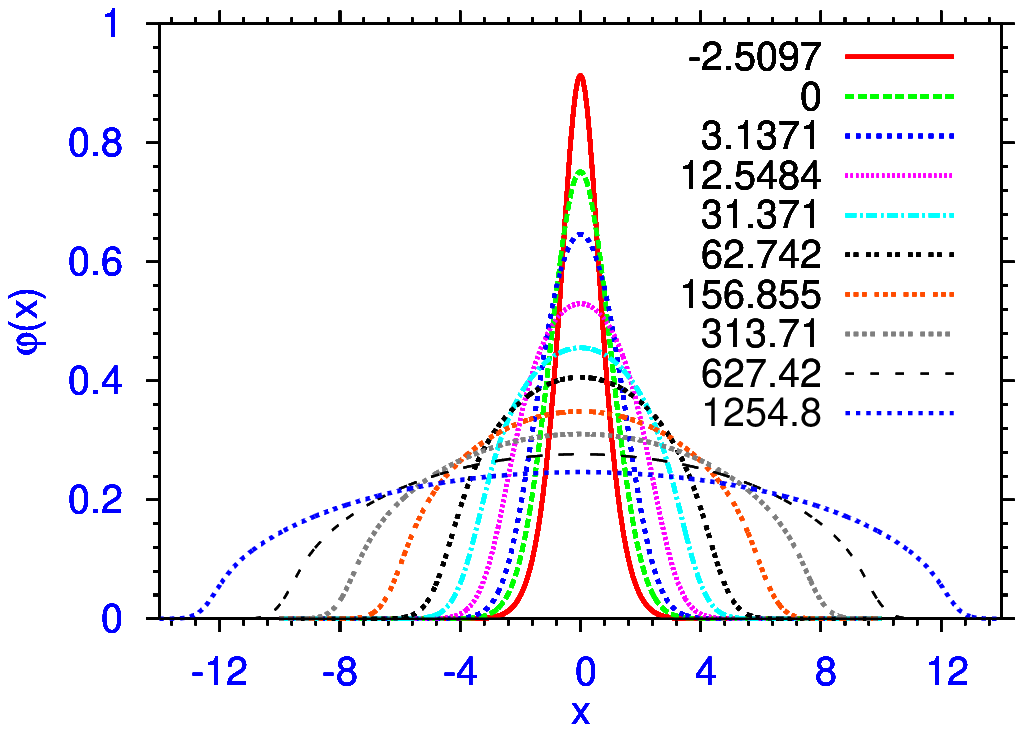}}
\end{center}
\caption{(Color online) Plot of wave function profile for the (a)
radially-symmetric 3D case [Eq. (\ref{sph2}), using imagtimesph.F,
OPTION 2]  and (b)  1D
case [Eq. (\ref{1d2}),  using imagtime1d.F, OPTION 2].  The
curves are
labeled by their respective nonlinearities as tabulated in Tables
\ref{table2} and \ref{table3}, respectively.}
\label{fig2}
\end{figure}

\begin{table}[!ht]
\begin{center}
\caption{Convergence of results for chemical potential $\mu$, rms size $r_{\mathrm{rms}}$ and
the wave function $\phi(0)$ at the center
in the Cartesian 2D
 case for a nonlinearity $\aleph 
=12.5484$ and anisotropy
$\kappa =1$ for different space steps $h\equiv$ DX = DY obtained from
the program imagtime2d.F
[using Eq. (\ref{2d2}), OPTION 2].}
\label{table4}
\begin{tabular}{|r|r|r|r|r|r|}
\hline
$\aleph 
$ & DX=DY  &  DT &
$\varphi(0)$ &
$r_{\mathrm{rms}}$ &
{$\mu$ } \\
\hline
  12.5484 & 0.10 & 0.01  &0.39189(2) &     1.30678(3) &   2.2559  \\
  12.5484 & 0.08 & 0.005  &0.39190(2) &     1.30685(3) &   2.2559  \\
  12.5484 & 0.06 & 0.003  &0.39190(2) &     1.30690(3) &   2.25582(3)
\\
  12.5484 & 0.04  & 0.001   & 0.39190(2)   &    1.30693(2) &
2.25579(2)
\\
  12.5484 & 0.02  & 0.0005  &0.39190(2)  &    1.30687(2) &
2.25583(2)
\\
  12.5484 & 0.015  & 0.0003  &0.39190(2)  &    1.30687(2) &
2.25583(2)
\\
  12.5484 & 0.01  & 0.0001  &0.39190(2)  &    1.30687(2) &
2.25583(2)
\\
\hline
\end{tabular}
\end{center}
\end{table}

In Figs. \ref{fig2} (a) and (b) we plot the wave function profiles for
the radially-symmetric 3D and  1D cases calculated using the
programs imagtimesph.F and imagtime1d.F, respectively,  for
different nonlinearities presented in Tables   \ref{table2}  and
\ref{table3}. As the nonlinearity is increased the system becomes more
repulsive and the wave function extends to a larger domain in space.

In Table \ref{table4}
we present  the results for the wave function at
center, rms
radius, and chemical potential for the Cartesian 2D case with
nonlinearity $\aleph  
$ = 12.5484 using the imaginary-time 
propagation
program imagtime2d.F.
We find that desired convergence is achieved with space step DX =
0.02. (Note that the converged result in this case is less accurate
than those in Tables \ref{table1}, \ref{table2} and \ref{table3} as we
have used a larger space step in order to keep the CPU time small. A
finer mesh will increase the accuracy requiring a larger CPU time.)

\begin{table}[!ht]
\begin{center}
\caption{The chemical potential $\mu$, rms size $r_{\mathrm{rms}}$,
and the wave function
$\varphi(0)$ at the center  for various
nonlinearities in the anisotropic  2D case obtained using the programs
imagtime2d.F [Eq. (\ref{2d2}), OPTION 2] and imagtimecir.F [Eq.
(\ref{cir2}), OPTION 2]. The
case $\kappa=1$
represents
circular symmetry and $\kappa\ne 1$ corresponds to anisotropy.
Table completed with
DX = DY $
\le 0.02$
and DT $=0.0001$ for imagtime2d.F and space step DR $\le 0.0025$ and DT $=0.00002$ for
imagtimecir.F. }
\label{table5}
\begin{tabular}{|r|r|r|r|r|r|r|r|r|r|r|}
\hline
$\kappa$ &
{$\aleph$  
} &
\multicolumn{1}{c}{} &
\multicolumn{1}{c}{$\varphi(0)$} &
      &
\multicolumn{1}{c}{}      &
\multicolumn{1}{c}{$r_{\mathrm{rms}}$} &
&
\multicolumn{1}{c}{}  &
\multicolumn{1}{c}{$\mu $}   &
 \\
\hline
&   & anisotropic &circular &\cite{Bao_Tang} &anisotropic &circular
&\cite{Bao_Tang}
&anisotropic &circular & \cite{Bao_Tang}\\
\hline
1&$-2.5097$ & 0.6754&0.67532(3) &  0.6754 & 0.87759(2)& 0.87758(1) &
0.8775  &
0.49978(3) &0.49978(1)
&
0.4997\\
1&       0 &  0.5642& 0.56419(1)  & 0.5642& 1.00000 & 1.00000
&1.0000 & 1.00000 & 1.000000 &
1.0000
\\
1&    3.1371 &0.4913& 0.49128(1) & 0.4913 &1.10513(2) & 1.10515(1)
&1.1051 & 1.42005(1)
&1.420054(3)  &
1.4200\\
1&   12.5484 & 0.3919& 0.39190(2) & 0.3919 &  1.30687(2)& 1.30686(1)
& 1.3068 &
2.25583(1) &
   2.255840(3)&
2.2558 \\
$\sqrt 2$&   12.5484 & 0.4267&  &  &  1.22054(2) & &  &
2.69607(1) &  &  \\
$ 2$&   12.5484 & 0.4633(1)&  &  &  1.17972(2)  &  & &
3.25488(1) &  &  \\
1&  62.742 & 0.2676&  0.26760(3) &0.2676  & 1.78817(2)&1.78816(1)
&1.7881  &  4.60982(1)
&  4.609831(3)  &
4.6098 \\
$1/\sqrt 2$&  62.742 & 0.2453&  &  & 1.99987(2)& &  &  3.88210(2) &
&
\\
1/ 2&  62.742 & 0.2249&  &  & 2.34157(2)&  & &  3.27923(2) & &    \\
1&313.71 &  0.1787 & 0.17872(3) &0.1787 &  2.60441(2) & 2.60441(1)
&2.6044 &
10.06825(3)&10.068262(5)
&10.068 \\
1&  627.42 & 0.1502 &0.15024(3) &0.1502 & 3.08453(2)  &3.08453(2)
&3.0845 &
14.18922(3)
& 14.189228(5)&
14.1892
\\
\hline
\end{tabular}
\end{center}
\end{table}

\begin{figure}[tbp] \begin{center}
{\includegraphics[width=.49\linewidth]{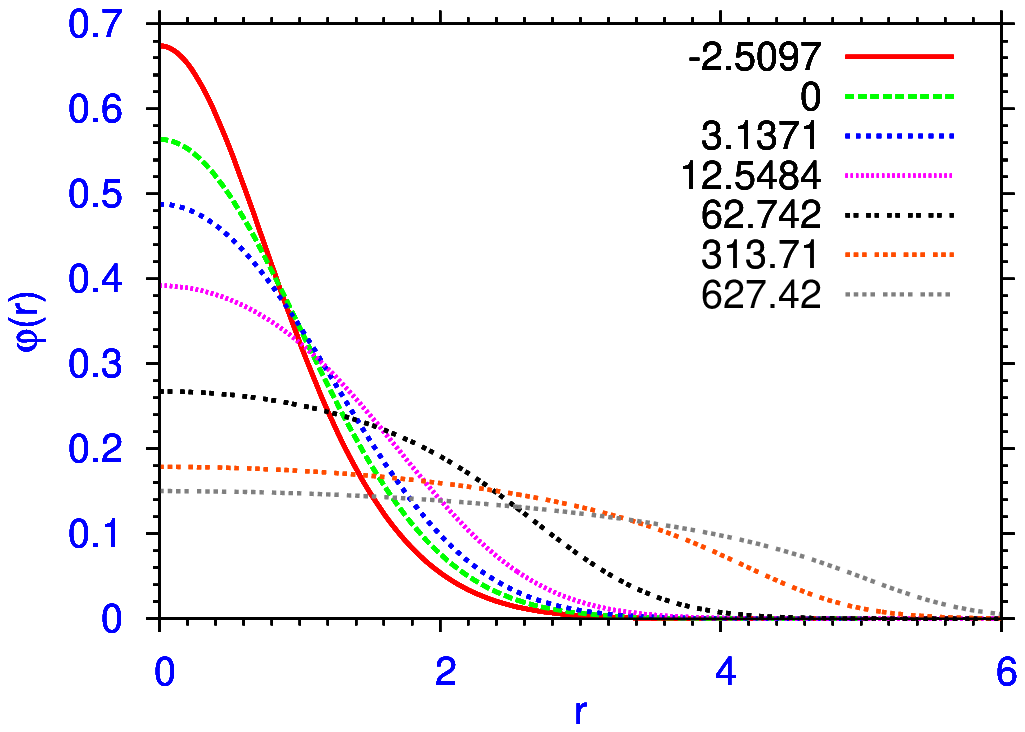}}
{\includegraphics[width=.49\linewidth]{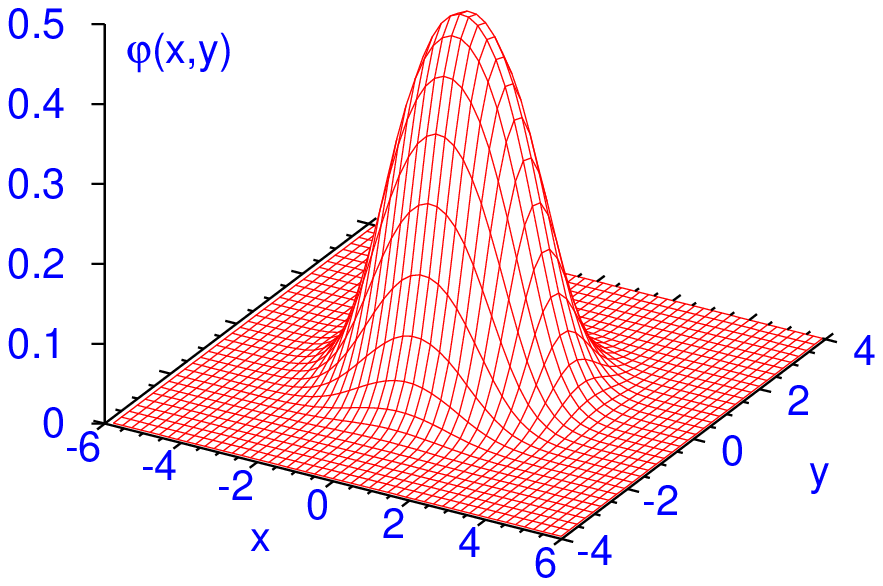}}
\end{center}
\caption{(Color online) Plot of wave function profile for the (a)
circularly-symmetric [Eq. (\ref{cir2}), OPTION 2]
and
(b)
anisotropic 2D cases [Eq. (\ref{2d2}), OPTION 2]
with nonlinearity $\aleph 
=12.5484$ and anisotropy
$\kappa=2$ using programs imagtimecir.F and imagtime2d.F, respectively.
Curves in (a) are labeled by the respective nonlinearities
as in Table  \ref{table5}.
}
\label{fig3}
\end{figure}

\begin{table}[!ht]
\begin{center}
\caption{The chemical potential $\mu$, rms sizes,
and the wave function
$\varphi(0)$ at the center  for various
nonlinearities in the axially-symmetric  3D case
for $^{87}$Rb atoms in an axial trap
with $\lambda=4,\kappa=1$
obtained using the
program
imagtimeaxial.F [Eq. (\ref{axi2}), OPTION 2]. The axial frequency is
taken as
$\omega_z=80\pi $ Hz,
$m(^{87}$Rb$)=1.44\times 10^{-25}$ kg,
$l=\sqrt{\hbar/m\omega_x}=0.3407\times 10^{-5}$ m, $a=5.1$ nm,
and the ratio between scattering and oscillator length
$4\pi a/l=0.01881$.
Table completed with DZ
$ \le 0.02$, D$\rho \le 0.02$ and
and DT $=0.00004$.}
\label{table6}
\begin{tabular}{|r|r|r|r|r|r|r|r|r|r|}
\hline
$N$        &
${\aleph}$  &
$\varphi(0)$ &
$\varphi(0)$\cite{Bao_Tang} &
$ \rho _{\mathrm{rms}}$  &
$\rho_{\mathrm{rms}}$\cite{Bao_Tang} &
$ z_{\mathrm{rms}}$ &
$  z_{\mathrm{rms}}$\cite{Bao_Tang} &
$\mu $  &$\mu $ \cite{Bao_Tang}\\
\hline
0& 0  &     0.5993(1) &  0.602  & 1.0000 & 1.000 &  0.3536 & 0.3539 &
3.0000  & 3.0000 \\
1000& 18.81  &       0.3813(2)  &0.3824 &1.3249  &1.325 &0.3805 & 0.3807
& 4.3611  & 4.362\\
5000& 94.05  &    0.2474     &0.2477 & 1.7742 &1.7742  &0.4212  &0.4214
& 6.6797  & 6.680 \\
10000& 188.1  &  0.2021       &0.2023 &2.0411 &2.041  & 0.4496  &0.4497
& 8.3671  & 8.367 \\
50000& 940.5  &   0.1247      &0.1248  &2.8424 & 2.842 &0.5531 & 0.5532
& 14.9487  &14.95  \\
100000& 1881  &   0.1011      &0.1012  &3.2758 & 3.276 &0.6173 & 0.6174
&  19.4751 & 19.47 \\
400000& 7524  &  0.0666       & 0.0666 &4.3408 &4.341  &0.7881 & 0.7881
& 33.4677  & 33.47 \\
800000& 15048  &      0.0540   & 0.0540 &4.9922 &4.992  & 0.8976 &
0.8976
& 44.0234  & 43.80 \\
\hline
\end{tabular}
\end{center}
\end{table}

\begin{figure}[tbp] \begin{center}
{\includegraphics[width=.49\linewidth]{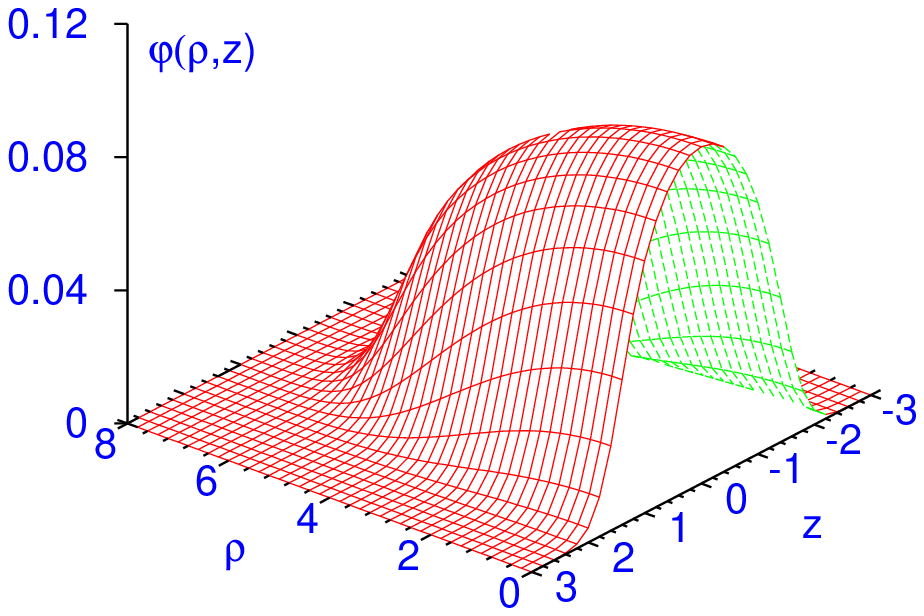}}
{\includegraphics[width=.49\linewidth]{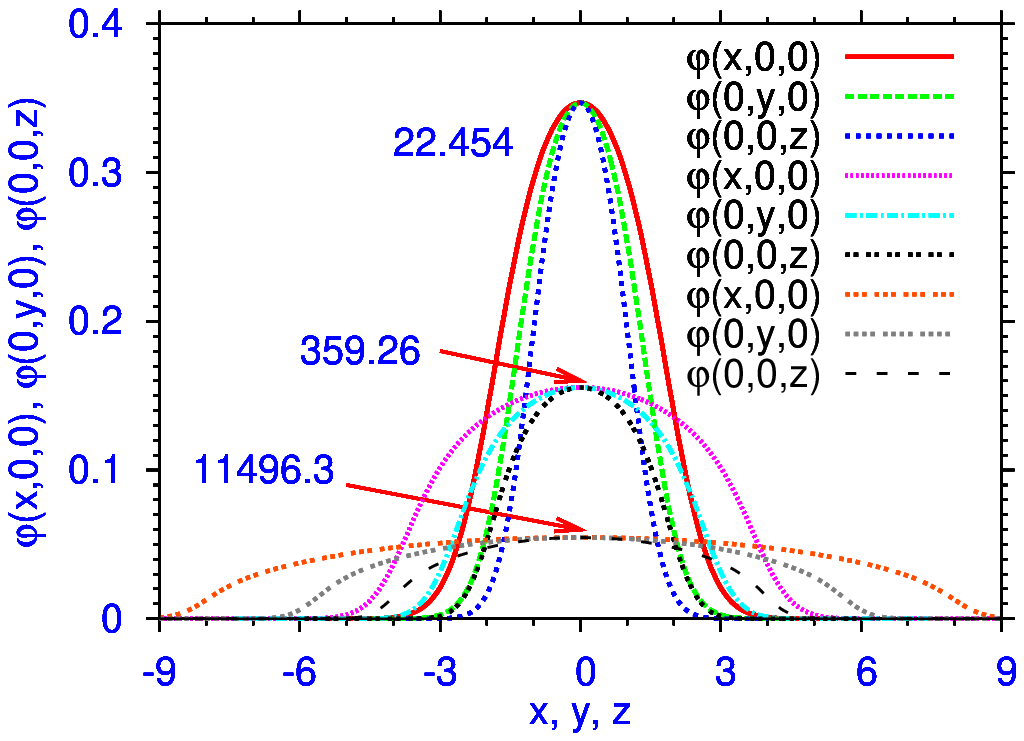}}
\end{center}
\caption{(Color online) Plot of wave function profile for the (a)
axially-symmetric 3D case with nonlinearity $\aleph 
=1881$
anisotropy
$\lambda=4, \kappa=1$  using imagtimeaxial.F [Eq. (\ref{axi2}), OPTION
2]
and (b) anisotropic 3D case with nonlinearity $\aleph 
=22.454,
359.26,$ and 11496.3 and anisotropy $\nu=1, \lambda =\sqrt 2$ and $\kappa =2$
 using imagtime3d.F [Eq. (\ref{ani2}), OPTION 2].
In the 3D case only the sections $\varphi(x,0,0), \varphi(0,y,0)$ and
$\varphi(0,0,z)$ of the wave functions are plotted.}
\label{fig4}
\end{figure}

\begin{table}[!ht]
\begin{center}
\caption{The chemical potential $\mu,$ rms sizes, and wave function 
$\varphi(0)$ at the center  for various number $N$ of
condensate of Na
atoms. The constants used are $m$(Na) $= 38.175\times 10^{-27}$ kg,
$a$(Na)
 = 2.75 nm. In all cases the input to numerical calculation was the
nonlinearity
coefficient $\aleph 
=4\pi Na/l$ shown below.
For the
spherically-symmetric
case, we solved the radially symmetric program imagtimesph.F [Eq.
(\ref{sph2}), OPTION 2],
(in
addition to the 3D
anisotropic program imagtime3d.F
setting equal frequencies in all three directions
with DX = DY =DZ = 0.05 and DT = 0.0004)
using \cite{Schneider_Feder,hau}
$\omega_0^S=87$ rad/s,
$
l=\sqrt{\hbar/m\omega_0^S}= 5.635$   $\mu$m, DR $\le 0.0025$ and
DT = 0.00002.
For the fully
anisotropic case we used the program imagtime3d.F [Eq.
(\ref{ani2}), OPTION 2] with \cite{Schneider_Feder,kozuma}
$\omega_x\equiv \omega_0 ^A=354 \pi$ rad/s,
$\omega_y=\sqrt 2
\omega_x,
\omega_z=2\omega_x$, $l=\sqrt{\hbar/m\omega_0^A}=1.576$ $\mu$m,
DX = DY =DZ = 0.05 and DT = 0.0004.}
\label{table7}
\begin{tabular}{|r|r|r|r|r|r|r|r|r|r|}
\hline
   & \multicolumn{1}{c}{} &
\multicolumn{1}{c}{Spherical} &  \multicolumn{1}{c}{}&{}
& \multicolumn{1}{c}{}
& \multicolumn{1}{c}{}
&\multicolumn{1}{c}{anisotropic}
& \multicolumn{1}{c}{}
& {}
\\
\hline
$N $
& {$\aleph $}  
& $\mu$(sph)&$\mu$(ani)  &$\mu $ 
\cite{Schneider_Feder}
& {$\aleph$}&$r_{\mathrm{rms}}$   &$\varphi(0)$&  $\mu$(ani) & $\mu $
\cite{Schneider_Feder} \\
\hline
0 &  0   &1.500000    & 1.5000 &1.500
& 0&1.0505 &0.5496 &
2.2071 &
2.207
\\
    1024 &  6.2798   &1.824546(1)& 1.8245   &  1.825 &
22.454&1.3211
&
0.3471& 3.5718&
3.572\\
    2048 &    12.5597 & 2.065406(1) & 2.0654 &2.065    &
44.907&
 1.4584
& 0.2888&
4.3446
&4.345 \\
    4096 &    25.1194 & 2.434526(1)  & 2.4345 &2.435  &
89.81
&1.6328
&0.2363 &5.4253
&5.425 \\
    8192 &   50.239  &  2.970180(1)  & 2.9702 &2.970  &
179.63&1.8460
&0.1919
&6.9042
&6.904 \\
   16384 &    100.477 & 3.719211(1)  & 3.7192 &3.719  &
359.26
&2.0999
&0.1555 &8.9003
&8.900\\
   32768 &   200.955 & 4.743445(2)  & 4.7434 &4.743  &
718.52&
2.3979
&0.1260 &11.5718
&11.572\\
   65536 &   401.91 &6.123751(2)  & 6.1238 &6.124  &
1437.03&2.7447
&0.1022
&  15.1284
&15.128
\\
  131072 &   803.82 & 7.970154(2)  & 7.9702 &7.970  &
2874.06 &  3.1460
&0.0829 & 19.8475& 19.847
\\
  262144 &  1607.64 & 10.426912(3)     & 10.4269  & 10.427  &
5748.13 &
3.6092&
0.0673&
26.0961&26.096
\\
  524288 &  3215.28 & 13.685486(3)  &13.6855 &13.685  &
11496.3
&4.1426
&0.0546 & 34.3590 &
34.358
\\
\hline
\end{tabular}
\end{center}
\end{table}

In Table \ref{table5} we present results for the wave function at the
center, rms size, and chemical potential for different nonlinearities
in the anisotropic  and circularly-symmetric 2D case calculated using
the imaginary-time routine
imagtime2d.F and imagtimecir.F,  respectively.  In the anisotropic
case we used
 space step 0.02 and time step 0.0001 and in the circularly-symmetric
case we used  space step 0.0025 and time step 0.00002. We also compare
these
results with those of Ref. \cite{Bao_Tang} and establish more accurate
results in the present numerical calculation.

In  Figs. \ref{fig3} (a) and (b) we plot the wave function profiles for
the circularly-symmetric and  anisotropic 2D  cases using the programs
imagtimecir.F
 and
imagtime2D.F, respectively.
In the anisotropic 2D case the anisotropy $\kappa =2$ and nonlinearity
$\aleph 
=12.4584$. Because of anisotropy the wave function in 
Fig. \ref{fig3}
(b)
is compressed in
the $y$ direction (note the different scales in $x$ and $y$ directions
of the plot.)

In Table \ref{table6} we present results for the wave function at the
center, rms sizes, and chemical potential for different nonlinearities
in the axially-symmetric 3D  case with $\kappa=1$ and $\lambda =4 $  calculated using
the imaginary-time routine
imagtimeaxial.F and  space step $D\rho=DZ=0.02$ and time step $DT=$ 0.0004. We
compare with the results of Ref. \cite{Bao_Tang} and establish more
accurate
results.

In Figs. \ref{fig4} (a) and (b) we plot the wave function profiles for
the axially-symmetric and fully anisotropic 3D cases using the programs
imagtimeaxial.F  and  imagtime3d.F, respectively.
In the axially-symmetric 3D case the anisotropy $\kappa=1,$ $\lambda =4$ and
nonlinearity $\aleph 
=1881$ were employed.
In the  fully anisotropic 3D case the anisotropies  $\nu=1$,  $\kappa
=\sqrt 2$, and
$\lambda =2$ and nonlinearities $ \aleph 
=22.454, 359.26$  and 
11496.3 are used. The effect of the anisotropy
is
explicit in the 3D case in generating different profiles of the
wave function along $x$, $y$, and $z$ directions for a fixed
nonlinearity.

Next we consider  the fully anisotropic case
in three dimensions and compare our results with those of Ref.
\cite{Schneider_Feder} calculated using the program imagtime3d.F. This
case mimics a realistic case of
experimental
interest with Na atoms. The completely anisotropic trap considered here
is time-orbiting potential (TOP) trap with angular frequencies in the
natural ratio  ($\omega_x,\omega_y,\omega_z)=\omega_0^A(1,\sqrt
2,2)$, with $\omega_0^A=354\pi $ rad/s. BEC in such a system has been
observed by Kozuma {\it et al.} \cite{kozuma}. We also consider
the
spherical potential with $\omega_0^S=87$ rad/s \cite{hau}. The s-wave
scattering
length of Na is taken as $a=52a_0\approx 2.75$ nm, with $a_0= 0.5292$
\AA \-  the Bohr
radius
\cite{Schneider_Feder}.

In Table \ref{table7} we exhibit the results for our calculations with
the fully anisotropic potential  together with those for the
spherically-symmetric potential in three dimensions. The results for the
spherically-symmetric potential are also calculated using the
one-dimensional
radially symmetric imaginary-time program imagtimesph.F
in addition to the fully
anisotropic program imagtime3d.F.  In the anisotropic case the
present results are consistent with those of Ref.
\cite{Schneider_Feder}.
In the spherical case the two sets of the present results (calculated
with
the  spherically-symmetric and anisotropic programs) as well as those of
Ref. \cite{Schneider_Feder} are consistent with each other.

\begin{figure}[tbp] \begin{center}
{\includegraphics[width=.45\linewidth]{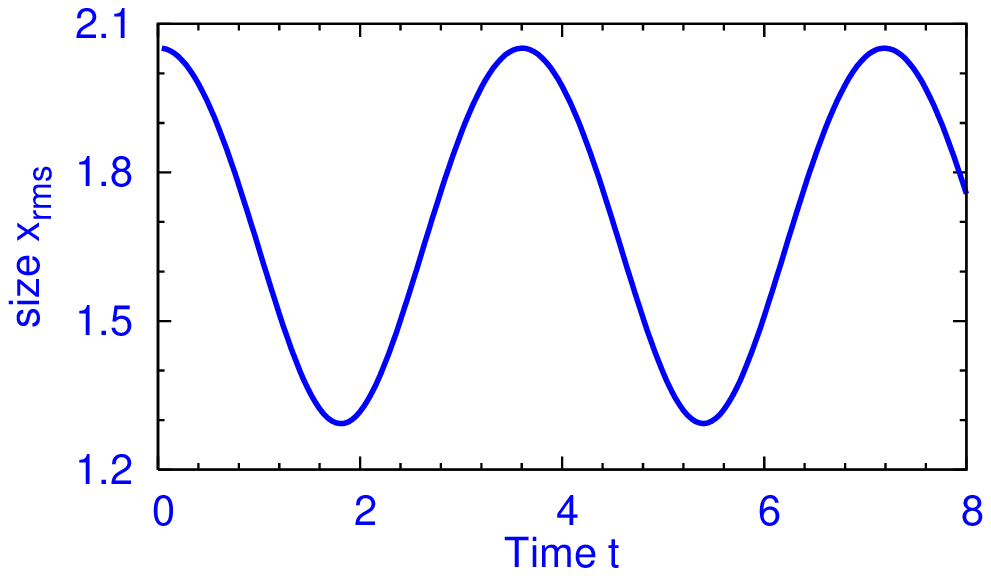}}
{\includegraphics[width=.45\linewidth]{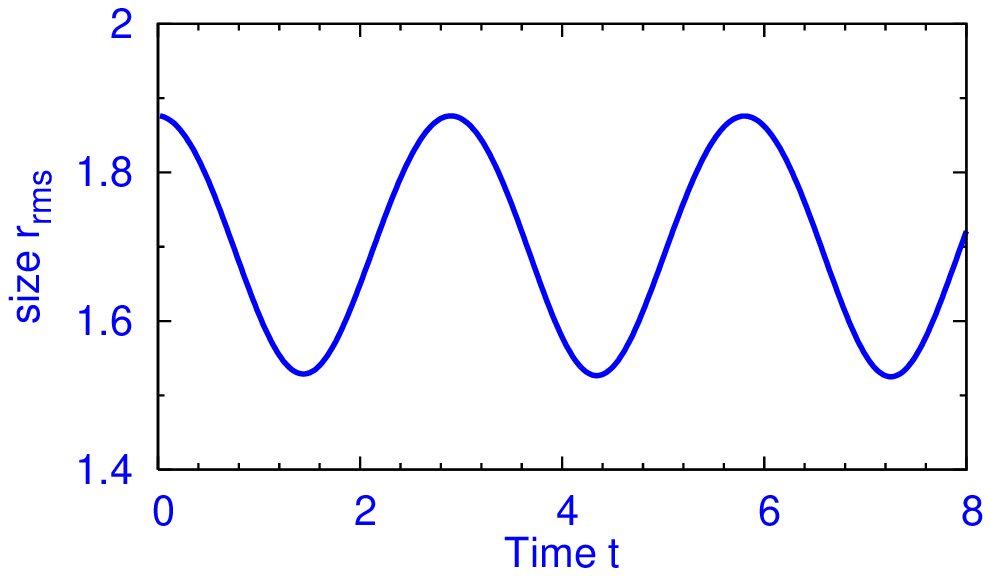}}
{\includegraphics[width=.45\linewidth]{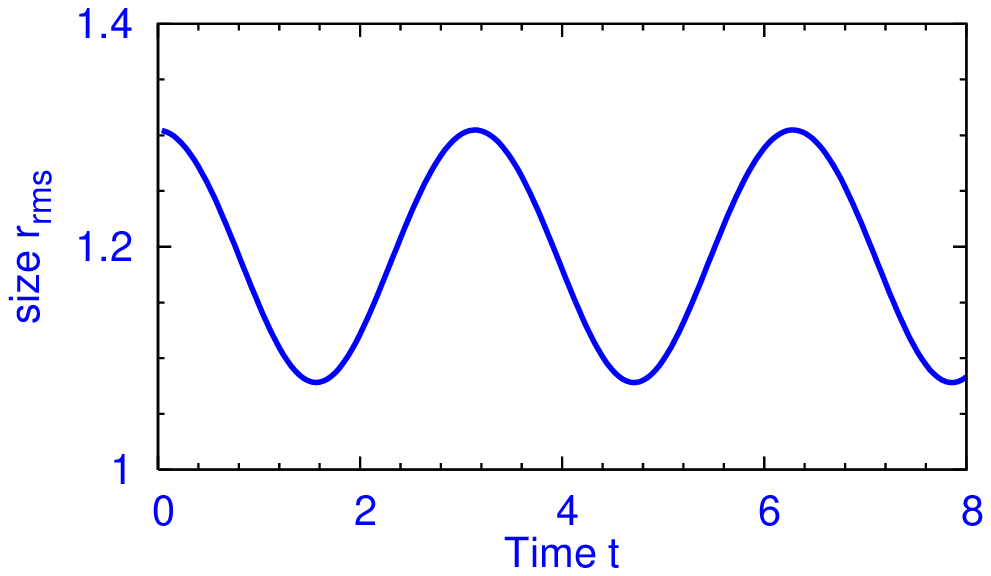}}
{\includegraphics[width=.45\linewidth]{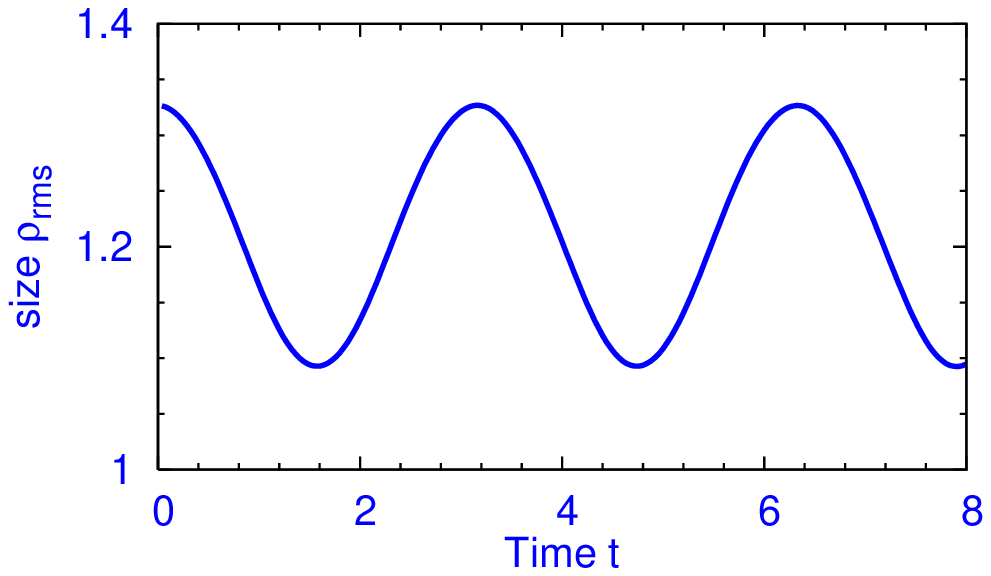}}
\end{center}
\caption{(Color online) Plot of rms size vs. time for non-stationary
oscillation of the system obtained by running the real-time programs for
(a) 1D  case (using program realtime1d.F with
$\aleph 
= 62.742$), (b)  spherically-symmetric case (using
program realtimesph.F with
$\aleph  
= 125.484$), (c) 2D circularly-symmetric case (using
program realtime2d.F
with $\aleph 
= 12.5484$ and $\kappa=1$),  and
(d) 3D axially-symmetric case (using program realtimeaxial.F  with
 $\aleph 
= 18.81$ and $\kappa=1, \lambda=4$).
 The oscillation is started during time evolution
by suddenly reducing the
nonlinearity $\aleph$  
to half after the formation of the 
stationary
condensate.}
\label{fig5}
\end{figure}

The input to our calculation is the nonlinearity coefficient $ \aleph$, 
which is related to the scattering length $a$, number of atoms $N$ and
harmonic oscillator length $l$ via $\aleph 
=4\pi a N/l$ in Eq.
(\ref{ani2}). We provide the nonlinearity values of our calculations.
Although the present results are in agreement with those of Ref.
\cite{Schneider_Feder}, a
very precise comparison of the two calculations
 is not to the point as Schneider and Feder  did not provide the
nonlinearity coefficient $\aleph 
$ used in their calculation.

\subsection{Non-stationary Oscillation}

The real-time propagation programs calculate the stationary states under
different trap symmetries. However, they are less efficient than the
imaginary-time propagation programs in this task requiring more CPU
time and producing less accurate results. However, unlike the
imaginary-time propagation programs, the real-time programs  can produce
time
evolution of non-stationary states also and next we present results of
such time evolution using the real-time propagation programs under
different trap symmetries.

In this subsection we present results for non-stationary oscillation
obtained with the use of the real-time programs. After the calculation
of the  stationary profile, the nonlinearity is suddenly reduced to half.
The wave function is no longer an eigenstate of the new nonlinear
equation. This sets the system into non-stationary oscillation which
continues for ever.  In Fig. \ref{fig5} we plot the rms size of the wave
function vs. time $t$ showing this oscillation using the output from
File 8  for (a)  1D case (using program realtime1d.F), (b)
radially-symmetric 3D case (using program realtimesph.F), (c)
Cartesian 2D case with anisotropy $\kappa=1$ (using program
realtime2d.F), and (d) 3D axially-symmetric case (using program 
realtimeaxial.F)  with  respective
nonlinearities $\aleph=$ 
62.742, 125.484, 12.5484, and 18.81.
The rms size at $t=0$
is the rms size of the stationary wave function obtained after NPAS time
iterations.

Because of transverse instability, the real-time program 
realtime3d.F in 3D does not lead to stable sinusoidal oscillation as in 
other
cases for a large change in nonlinearity (nonlinearity reduced to half 
of 
its initial value) as shown in Fig. \ref{fig5}.
Only for small perturbation a sinusoidal oscillation is observed.
However, we do not present a systematic study of such oscillation.

\section{Summary and Conclusion}
\label{SUM}

In this paper we describe a split-step method for the numerical solution
of the time-dependent nonlinear GP equation under the action of a
general anisotropic 3D trap using real- and imaginary-time propagation.
Similar methods for 1D and anisotropic 2D traps are also described. The
time propagation is carried with an initial input. The full Hamiltonian
is split into several spatial derivative and a non-derivative parts. The
spatial derivative parts are treated by the Crank-Nicolson method.
Different spatial derivative and non-derivative parts are dealt in
independent steps. This, so called  split-step, method leads to highly
stable and accurate results.

We considered two types of time iterations $-$ real-time propagation and
imaginary-time propagation. In the real-time propagation
time evolution  is performed with the original complex equation. The
numerical algorithm in this case requires the use of complex variable
but produces solution of non-stationary problems. In the imaginary-time
propagation, the time variable is replaced by i ($=\sqrt{-1})$ times a
new time variable, consequently the GP equation becomes real. The
numerical solution of this equation  can no longer yield the solution of
non-stationary problems; but yields very accurate solution of stationary 
ground state 
problems only, requiring  much smaller CPU time.

We provide the numerical algorithm in detail in 1D, 2D, and 3D for real-
and imaginary-time propagations.  We consider six different harmonic 
oscillator trap
symmetries, e.g., a  1D trap, a circularly-symmetric 2D trap, a
radially-symmetric 3D trap, an anisotropic trap in 2D, an
axially-symmetric 3D trap, and an anisotropic trap in 3D. Each of these
cases are treated with real- and imaginary-time propagation algorithms
resulting in twelve different Fortran 77 programs supplied. We use the
imaginary-time propagation programs to provide results for different
stationary properties of the condensate (chemical potential, rms size,
etc) in 1D, 2D, 3D, for different nonlinearities $\aleph $ 
and compare
with previously obtained results \cite{Bao_Tang,Schneider_Feder}. In
addition we study a non-stationary oscillation initiated by suddenly
altering the nonlinearity to half its initial value on these preformed
condensates using the real-time propagation programs. In addition six
Fortran 90/95 programs are supplied in the case of two and three space
variables.

Although the present programs are valid for   the standard GP equation
with cubic nonlinearity in a harmonic potential, they can be easily
adopted for other types of
bosonic \cite{tg} or fermionic equations \cite{ska1,ska2} with different
nonlinearities and under
different types of potentials. To change the potential one should change
the variable V in the subroutine INITIALIZE and the change in the
nonlinearity can be performed in the subroutine NONLIN.

\ack

We thank Dr. A. Gammal for helpful comments regarding the
solution of the GP equation in the circularly-symmetric and
axially-symmetric cases.
We thank Prof. W. Bao for the hospitality at the National University of
Singapore when this project was started. The research was partially
supported by the CNPq and FAPESP of Brazil, and the Institute for
Mathematical Sciences of the National University of Singapore.
PM thanks  the Third World Academy of Sciences (TWAS-UNESCO
Associateship at the Center of Excellence in the South), and Department
of Science and Technology, Government of India for partial
support.


\begin{thebibliography}{99}

\bibitem{review}
F. Dalfovo, S. Giorgini, L. P. Pitaevskii, S. Stringari, Theory of
{B}ose-{E}instein condensation in trapped gases, Rev. Mod. Phys. 71
(1999) 463-512;

A. J. Leggett,
Bose-Einstein condensation in the alkali gases: Some fundamental
concepts,
 Rev. Mod. Phys.
 73 (2001) 307-356.

\bibitem{books} L. Pitaevskii, S. Stringari,
Bose-Einstein Condensation, Clarendon Press, Oxford and New York, 2003;


 C. J. Pethick, H. Smith,
 Bose-Einstein Condensation in Dilute Gases, Cambridge University Press,
Cambridge, 2002.  	
	

\bibitem{Tiwari_Shukla}
R. P. Tiwari, A. Shukla, {A basis-set based {F}ortran program to solve
the {G}ross-{P}itaevskii equation for dilute {B}ose gases in harmonic
and anharmonic traps}, Comp. Phys. Commun. 174 (2006) 966-982.

\bibitem{Bao_Tang}
W. Bao, W. Tang, {Ground-state solution of {B}ose-{E}instein
condensate by directly minimizing the energy functional}, J. Comput.
Phys. 187 (2003) 230-254.

\bibitem{Schneider_Feder}
B.~I.~Schneider,  D.~L.~Feder, {Numerical approach to the ground and
excited state of a {B}ose-{E}instein condensed gas confined in a
completely anisotropic trap}, Phys. Rev. A 59 (1999) 2232-2242.

\bibitem{chio1}
M. L. Chiofalo, S. Succi,  M. P. Tosi,  Ground state of trapped
interacting Bose-Einstein condensates by an explicit imaginary-time
algorithm, Phys. Rev. E  62 (2000) 7438-7444.

\bibitem{chio2}
M. M. Cerimele, M. L. Chiofalo, F. Pistella, S. Succi, M. P. Tosi,
Numerical solution of the Gross-Pitaevskii equation using an explicit
finite-difference scheme: An application to trapped Bose-Einstein
condensates, Phys. Rev. E  62 (2000) 1382-1389.

\bibitem{chang}
S. L. Chang, C. S. Chien, B. W. Jeng, Computing wave functions of
nonlinear Schr\"odinger equations: A time-independent approach, J.
Comput. Phys. 226 (2007) 104-130.

\bibitem{num1}S. Palpacelli, S. Succi, R. Spigler, Ground-state computation of
Bose-Einstein condensates by an imaginary-time quantum lattice
Boltzmann scheme, Phys. Rev. E 76 (2007)  036712.

\bibitem{num2}
M. Javidi, A. Golbabai, Numerical studies on nonlinear Schr\"odinger
equations by spectral collocation method with preconditioning, J.
Math. Analysis and Applications  333 (2007) 1119-1127.

\bibitem{num3}
H. Q. Wang, A time-splitting spectral method for coupled
Gross-Pitaevskii equations with applications to rotating Bose-Einstein
condensates, J.  Comput and Applied Maths. 205 (2007) 88-104.

\bibitem{num4}
M. Brtka, A. Gammal, L. Tomio, Relaxation algorithm to hyperbolic
states in Gross-Pitaevskii equation, Phys. Lett. A 359 (2006) 339-344.


\bibitem{num5}
W. Hua, X. H. Liu, P. H. Ding, Numerical solution for the
Gross-Pitaevskii equation, J. Math. Chem. 40 (2006) 243-255.

\bibitem{num6}
Z. L. Xu, H.  Han, Absorbing boundary conditions for nonlinear
Schr\"odinger equations, Phys. Rev. E 74 (2006)  037704.


\bibitem{num7}
J. Javanainen, J. Ruostekoski, Symbolic calculation in development of
algorithms: split-step methods for the Gross-Pitaevskii equation, J.
Phys. A 39 (2006) L179-L184.

\bibitem{num8}
B. I. Schneider, L. A. Collins, S. X. Hu, Parallel solver for the
time-dependent linear and nonlinear Schr\"odinger equation, Phys. Rev.
E  73 (2006) 036708.

\bibitem{num9}
S.  Succi, F. Toschi, M. P. Tosi, et al., Bose-Einstein condensates
and the numerical solution of the Gross-Pitaevskii equation, Computing
Sci. Engineering 7 (2005) 48-57.

\bibitem{num10}
S. M. Chang, Y. C. Kuo, W. W. Lin, et al. A continuation
BSOR-Lanczos-Galerkin method for positive bound states of a
multi-component Bose-Einstein condensate, J. Comput. Phys. 210 (2005)
439-458.

\bibitem{num11}
W. Z. Bao, J. Shen, Fourth-order time-splitting Laguerre-Hermite
pseudospectral method for Bose-Einstein condensates,
Siam J.  Sci.
Computing 26 (2005) 2010-2028.

\bibitem{num12}
S. M. Chang, W. W. Lin, S. F. Shieh, Gauss-Seidel-type methods for
energy states of a multi-component Bose-Einstein condensate, J.
Comput.  Phys.  202 (2005) 367-390.

\bibitem{num13}
W. Z.  Bao, Q. Du, Computing the ground state solution of
Bose-Einstein condensates by a normalized gradient flow, Siam J.
Sci.  Computing 25 (2004) 1674-1697.

\bibitem{num14}
M. C.  	Lai, C. Y. Huang, T. S. Lin, A simple Dufort-Frankel-type
scheme for the Gross-Pitaevskii equation of Bose-Einstein condensates
on different geometries, Numerical Methods  Partial Diff. Eqs. 20
(2004) 624-638.

\bibitem{num15}
W. Z.  Bao, Ground states and dynamics of multicomponent Bose-Einstein
condensates, Multiscale Modeling  Simulation 2 (2004) 210-236.

\bibitem{num16}A. 	Zhou,
An analysis of finite-dimensional approximations for the ground state
solution of Bose-Einstein condensates,
Nonlinearity 17 (2004) 541-550.

\bibitem{num17}Y. S. 	Choi, J. Javanainen, I. Koltracht, et al.,
A fast algorithm for the solution of the time-independent
Gross-Pitaevskii equation, J. Comput. Phys.
190 (2003) 1-21.

\bibitem{num18}
P. Muruganandam, S. K. Adhikari, Bose-Einstein condensation dynamics
in three dimensions by the pseudospectral and finite-difference
methods, J. Phys. B 36 (2003) 2501-2513.


\bibitem{num19}
W. Z. Bao, S. Jin, P. A. Markowich, On time-splitting spectral
approximations
for the Schr\"odinger equation in the semiclassical regime,
J. Comput.
Phys. 175 (2002) 487-524.



\bibitem{num20}
W. Z. Bao, D. Jaksch, P. A. Markowich, Numerical solution of the
Gross-Pitaevskii equation for Bose-Einstein condensation, J. Comput.
Phys. 187 (2003) 318-342.

\bibitem{num21}
P. Vignolo, M. L. Chiofalo, M. P. Tosi, et al., Explicit
finite-difference and particle method for the dynamics of mixed
Bose-condensate and cold-atom clouds,  J. Comput. Phys. 182
(2002) 368-391.

\bibitem{num22}
S. K. Adhikari, P. Muruganandam, Bose-Einstein condensation dynamics
from the numerical solution of the Gross-Pitaevskii equation, J. Phys.
B 35 (2002) 2831-2843.

\bibitem{num23}
M. M.  Cerimele, M. L. Chiofalo, F. Pistella, Numerical solution of
the stationary Gross-Pitaevskii equation: tests of a combined
imaginary-time-marching technique with splitting, Nonlinear
Analysis-Theory Methods Applications 47 (2001) 3345-3356.

\bibitem{num24}W. Z. Bao, S. Jin, P. A. Markowich,
Numerical study of time-splitting spectral discretizations of nonlinear
Schr\"odinger equations in the semiclassical regimes,
Siam J.  Sci.
Computing 25 (2003) 27-64.

\bibitem{num25}
S. K. Adhikari, Numerical solution of the two-dimensional
Gross-Pitaevskii equation for trapped interacting atoms, Phys. Lett. A
265 (2000) 91-96.

\bibitem{num26}
A. Gammal, T. Frederico, L. Tomio, Improved numerical approach for the
time-independent Gross-Pitaevskii nonlinear Schr\"odinger equation,
Phys. Rev. E 60 (1999) 2421-2424.

\bibitem{num27}
A. L. Fetter, Variational study of dilute bose condensate in a
harmonic trap, J. Low Temp. Phys. 106 (1997) 643-652.

\bibitem{num28}
R. J. Dodd, Approximate solutions of the nonlinear Schr\"odinger
equation for ground and excited states of Bose-Einstein condensates,
J. Res.  National Inst.  Standards Tech. 101 (1996) 545-552.

\bibitem{num29}
S. K. Adhikari, Numerical study of the spherically symmetric
Gross-Pitaevskii equation in two space dimensions, Phys. Rev. E 62
(2000) 2937-2944.

\bibitem{num30}
S. K. Adhikari, Numerical study of the coupled time-dependent
Gross-Pitaevskii equation: Application to Bose-Einstein condensation,
Phys. Rev. E 63 (2001) 056704.

\bibitem{num31}
F. Dalfovo, S. Stringari, Bosons in anisotropic traps: Ground state
and vortices, Phys. Rev. A  53 (1996) 2477-2485.

\bibitem{num32}
Q. X.  Yuan, G. H. Ding, Computing ground state solution of
Bose-Einstein condensates trapped in one-dimensional harmonic
potential, Communications Theor. Phys.  46 (2006) 873-878.


\bibitem{num33}M. J. Holland, J. Cooper, Expansion of a Bose-Einstein
condensate in a harmonic potential,
Phys. Rev. A 53 (1996) R1954-R1957.

\bibitem{num34}M. M.  Cerimele, F. Pistella, S. Succi,
Particle-inspired scheme for the Gross-Pitaevski equation: An
application to Bose--Einstein condensation,
Comput. Phys. Commun. 129 (2000) 82-90.

\bibitem{num35}S. Palpacelli, S. Succi,
Quantum lattice Boltzmann simulation of expanding Bose-Einstein
condensates in random potentials,
Phys. Rev. E 77 (2008) 066708.

\bibitem{aq}
A. Aftalion, Q. Du, Vortices in a rotating Bose-Einstein condensate:
Critical angular velocities and energy diagrams in the Thomas-Fermi
regime, Phys. Rev. A 64 (2001) 063603.

\bibitem{xyz1}
  	L. Lehtovaara, 	 
J. Toivanen,
J. Eloranta, Solution of time-independent Schr\"odinger equation by the 
imaginary time propagation method, J. Comput. Phys. 
{221} (2007) 148-157. 

\bibitem{burnett}\
M. Edwards, K. Burnett, Numerical solution of the nonlinear
Schr\"{o}dinger equation for small samples of trapped neutral atoms,
Phys. Rev. A 51 (1995) 1382-1386.

\bibitem{holland}
P.~A.~Ruprecht, M.~J.~Holland,  K.~Burnett, M.~Edwards, Time-dependent
solution of the nonlinear {S}chr\"{o}dinger equation for
{B}ose-condensed trapped neutral atoms, Phys. Rev. A 51 (1995)
4704-4711.

\bibitem{baer}
R. Baer, Accurate and efficient evolution of  nonlinear
Schr\"{o}dinger equations,
Phys. Rev. A 62 (2000) 063810.


\bibitem{koonin}
S. E. Koonin, D. C. Meredith, Computational Physics: Fortran version,
Addison-Wesley, Reading, 1990.

\bibitem{ames}
W. F. Ames, Numerical Methods for Partial Differential Equations, 3rd
Ed, Academic Press, New York, 1992.

\bibitem{dtray}R. Dautray, J. L. Lions, { Mathematical Analysis
and Numerical Methods for Science and Technology, vol 6,} Springer
Verlag, Berlin, 1993, Chapter XX, Sec. 2.

\bibitem{ska5}
 S. K. Adhikari,
{ Mean-field description of collapsing and exploding
Bose-Einstein condensates},
Phys. Rev. A 66 (2002) 013611;

 S. K. Adhikari, { Dynamics of a collapsing and exploding Bose-Einstein condensed vortex
state}
Phys. Rev. A 66 (2002) 043601.



\bibitem{ska1}S. K. Adhikari, {  Nonlinear Schr\"odinger equation for
a superfluid Fermi gas in the BCS-BEC crossover},
Phys. Rev. A 77 (2008) 045602.

\bibitem{ska2}S. K. Adhikari, {    Superfluid Fermi-Fermi mixture:
Phase diagram, stability, and soliton formation},
Phys. Rev. A 76 (2007) 053609.

\bibitem{tg}  L. Salasnich, S. K. Adhikari, F. Toigo,
{  Self-bound droplet of Bose and Fermi atoms in one
dimension: Collective properties in mean-field and Tonks-Girardeau
regimes},
Phys. Rev. A 75 (2007) 023616;

M. Girardeau, Relationship between systems of impenetrable bosons and
fermions in one dimension,
J. Math. Phys. 1 (1960) 516-523

M. D. Girardeau, Permutation symmetry of many-particle wave functions,
Phys. Rev. 139 (1965) B500-B508.




\bibitem{ska3}
S. K. Adhikari,
{ Fermionic bright soliton in a boson-fermion mixture},
Phys. Rev. A 72 (2005)  053608;

S. K. Adhikari,
{ Stabilization of bright solitons and vortex solitons in a
trapless
three-dimensional Bose-Einstein condensate by temporal modulation of the
scattering length},
Phys. Rev. A 69 (2004) 063613;

S. K. Adhikari,
{ Mean-field model of interaction between bright vortex solitons in
Bose-Einstein condensates},
New J. Phys. 5 (2003) 137;

S. Rajendran,  P. Muruganandam, M.  Lakshmanan,
Non-stationary excitations in Bose-Einstein condensates under the action
of periodically varying scattering length with time-dependent
frequencies, Physica D 227 (2007) 1-7.

\bibitem{ska4}  S. K. Adhikari,
{ Free expansion of attractive and repulsive Bose-Einstein
condensed vortex states},
Phys. Rev. A 65 (2002) 033616.




\bibitem{kozuma}  M. Kozuma, L. Deng, E. W. Hagley, J. Wen,
R. Lutwak,
K. Helmerson, S. L. Rolston, W. D. Phillips,
{ Coherent Splitting of Bose-Einstein Condensed Atoms with Optically
Induced Bragg Diffraction},
Phys. Rev. Lett.  82 (1999) 871-880.



\bibitem{hau}
L. V. Hau, B. D. Busch, C. Liu, Z. Dutton,
M. M. Burns, J. A. Golovchenko,
{ Near-resonant spatial images of confined Bose-Einstein condensates
in a 4-Dee magnetic bottle},
Phys. Rev. A 58 (1998) R54-R57.




\end{thebibliography}

\end{document}